\documentclass[twocolumns,10pt]{IEEEtran}

\usepackage{amssymb,amsmath,amsfonts,amsthm}
\usepackage{mathrsfs}
\usepackage{cite}
\usepackage{array}
\usepackage{tabularx}
\usepackage{color}
\usepackage{mathtools}
\usepackage{subfigure}
\usepackage{graphicx}
\usepackage{bbm}

\newtheorem{definition}{Definition}

\newtheorem{lemma}{Lemma}

\newtheorem{remark}{Remark}
\newcommand{\argmax}{\mathop{\rm arg~max}\limits}

\IEEEoverridecommandlockouts

\begin{document}
\title{Cooperative Caching and Transmission Design in Cluster-Centric Small Cell Networks}


\author{Zheng~Chen,~\IEEEmembership{Student Member,~IEEE,} Jemin~Lee,~\IEEEmembership{Member,~IEEE,} Tony~Q.~S.~Quek,~\IEEEmembership{Senior Member,~IEEE,}
	and~Marios~Kountouris,~\IEEEmembership{Senior Member,~IEEE}
	\thanks{
	A part of this paper has been submitted to the 2016 IEEE International Conference on Communications.}
	\thanks{Z. Chen is with the Laboratoire de Signaux et Syst\`{e}mes (L2S, UMR8506),
		CentraleSup\'{e}lec - CNRS - Universit\'{e} Paris-Sud,
		Gif-sur-Yvette, France. (email: zheng.chen@centralesupelec.fr)}
	\thanks{J. Lee and T. Q. S. Quek are with Singapore University of Technology and Design (SUTD). (email: \{jemin\_lee, tonyquek\}@sutd.edu.sg) }
	\thanks{M. Kountouris is with the Mathematical and Algorithmic Sciences Lab, France Research Center, Huawei Technologies Co. Ltd. (email: marios.kountouris@huawei.com) }
}

\maketitle

\begin{abstract}	
Wireless content caching in small cell networks (SCNs) has recently been considered as an efficient way to reduce the traffic and the energy consumption of the backhaul in emerging heterogeneous cellular networks (HetNets).
In this paper, we consider a cluster-centric SCN with combined design of cooperative caching and transmission policy. Small base stations (SBSs) are grouped into disjoint clusters, in which in-cluster cache space is utilized as an entity. We propose a combined caching scheme where part of the available cache space is reserved for caching the most popular content in every SBS, while the remaining is used for cooperatively caching different partitions of the less popular content in different SBSs, as a means to increase local content diversity. Depending on the availability and placement of the requested content, coordinated multipoint (CoMP) technique with either joint transmission (JT) or parallel transmission (PT) is used to deliver content to the served user. Using Poisson point process (PPP) for the SBS location distribution and a hexagonal grid model for the clusters, we provide analytical results on the successful content delivery probability of both transmission schemes for a user located at the cluster center. Our analysis shows an inherent tradeoff between transmission diversity and content diversity in our combined caching-transmission design. We also study optimal cache space assignment for two objective functions: maximization of the cache service performance and the energy efficiency. Simulation results show that the proposed scheme achieves performance gain by leveraging cache-level and signal-level cooperation and adapting to the network environment and user QoS requirements.
\end{abstract}

	
\section{Introduction}
Current cellular networks are under continuously increasing pressure mainly due to the exponentially growing wireless data traffic and the pressing demand for capacity boosting and enhanced uniform coverage. Network densification through deployment of heterogeneous infrastructure, e.g., pico base stations and femto access points (FAPs), is envisioned as a promising solution to improve area spectral efficiency and network coverage. Nevertheless, in dense small cell network (SCNs) deployment, backhaul availability and capacity may become the performance and cost bottleneck. 
Cache-enabled SCNs have been proposed as a potential solution for the backhaul bottleneck  ~\cite{caching_edge, femtocaching, femtocaching2013, clustering}. The main idea is to introduce cache capabilities at small base stations (SBSs) and then prefetch content during off-peak hours before being requested locally by the end users. Caching in wireless networks also exploits the high degree of asynchronous content reuse caused by information-centric applications, such as video-on-demand (VoD), social networks, and content sharing. When end users request for some popular content being already cached in the local SBSs, the service latency is largely reduced since there is no need to pass through the backhaul to retrieve the content from remote servers. The improved energy efficiency is also an important benefit of small cell caching mainly due to the fact that repeated transmissions of the same content from the core network to local SBSs are avoided. 

Different from existing cache replacement algorithms in Internet caching, proactive caching in wireless networks requires caching content closer to potential users so as to increase the hit rate and the probability of successful delivery.  In the literature of proactive caching in SCN, the selection of content to be cached is usually based on some known popularity information on the content. Most prior work use ``homogeneous" caching strategies, meaning that different SBSs either cache the same popular content or cache with the same probabilistic placement policy. 
The conventional ``cache the most popular content everywhere'' strategy, which corresponds to the Least Frequently Used (LFU) replacement policy in Internet caching, gives optimal performance with non-overlapping SBS coverage or with isolated caches. When SBSs have overlapping coverage areas, users have more than one potential serving SBSs. The cache hit ratio can be improved by adopting an optimal probabilistic placement policy to increase content diversity in the caches of potential serving SBSs \cite{optimalcaching, random_caching}. Nevertheless, when users can simultaneously be served by multiple SBSs, with cooperation enabled not only in cache space (i.e., cache-level cooperation),  but also in the physical layer for content delivery (i.e., signal-level cooperation), the optimal cache placement design is expected to be different from the single serving SBS case.

\subsection{Related Work}
In heterogeneous networks, coexistence between SBSs and conventional macro base stations causes additional inter-cell interference when spectrum resources are shared \cite{optimization, interference_coordination}. Coordinated multipoint (CoMP) techniques have been proposed to mitigate inter-cell interference and increase network coverage and cell-edge throughput by allowing geographically separated BSs to communicate cooperatively \cite{comp}.  There are many types of CoMP techniques depending on how the cooperation is performed, e.g., coordinated scheduling/beamforming (CS/CB) and joint processing (JP). 
CoMP joint transmission for downlink heterogeneous cellular networks with randomly located users is studied in \cite{JC}, where analytical expressions for the coverage probability and the diversity gain are derived using tools from stochastic geometry. 

Recent studies in wireless caching with CoMP techniques provide new perspectives on the benefits of caching to achieve physical layer (PHY) cooperation gain. \cite{liu2015asymptotic} proposes a PHY caching scheme called cache-induced dual-layer CoMP, and studies asymptotic scaling laws of wireless ad hoc network with such scheme.
Considering cooperative transmission via caching helpers, \cite{cooperative_transmission} investigates the optimal caching placement as a means to balance diversity and cooperation gain.   
In addition to signal-level cooperative transmission, cache-level cooperation in SCN can be realized by considering the caching capabilities of multiple SBSs as an entity and selectively cache different contents in different SBSs. By doing so, the cache hit probability of user requests can be improved because of increased content diversity. However, the probability to find a requested file cached in a SBS is no longer the same for all the SBSs, requiring local centralized control for cache placement decisions. The idea of cache-level cooperation has been discussed in the literature in different scenarios. In \cite{small_cell_coop}, the authors study small cell cooperation with threshold-based caching method to combine the advantages of distributed caching and PHY layer cooperative transmission. Backhaul-aware caching placement strategy for a group of cooperative BSs is studied in \cite{backhaul_aware} by solving an optimization problem to minimize the average download delay. 
Nevertheless, none of the existing works provide efficient solutions for the cache utilization policy in cooperative SCNs without relying on iterative algorithms.

\subsection{Contributions}
This paper proposes a cluster-centric SCN with combined design of the caching policy and cooperative transmission in order to optimally balance both transmission and content diversity. The overall cache space within a cluster is arranged by central controllers so as to either distribute the same popular content in every SBS or store different partitions of the less popular content in different SBSs, ensuring that all partitions of cached content can be found inside the cluster. Every SBS has the same proportion of cache space assigned for the most popular content (MPC), while the remaining is used to achieve largest content diversity (LCD). Depending on whether the content is cached using either MPC or LCD strategy, we use two transmission (delivery) schemes when a cache hit happens, namely coordinated joint transmission (JT) and parallel transmission (PT) with successive interference cancellation (SIC). 

For our cooperative caching-transmission design, we derive the successful content delivery probability (SCDP) of both CoMP transmission schemes for a user located at the cluster center. The cache hit probability is given as a function of the proportion of cache space assigned for MPC caching. We show that there exists an inherent tradeoff between transmission diversity and content diversity in such cluster-centric cooperative SCNs. We then solve two optimization problems for the optimal MPC caching proportion considering the two following objective functions: maximization of the cache service probability and of the energy efficiency. The optimal solution can be applied directly at the central controllers, adapting to the network environment, user quality-of-service (QoS) requirement and content popularity information. 

This paper is organized as follows. We present the network model and cooperation schemes in Section \ref{section_2}. In Section \ref{SCDP} we define the SCDP as the main performance metric and give analytical results for JT and PT transmission schemes. Based on numerical results of the SCDP and cache hit probability, in Section \ref{optimal caching} we show the tradeoff between transmission diversity and content diversity. We then pose two optimization problems for the optimal cache space assignment. Simulation results are presented in Section \ref{simulation} and Section \ref{conclusion} concludes the paper.

\section{Network Model}
\label{section_2}

\begin{figure}
	\centering
	\includegraphics[scale=0.6]{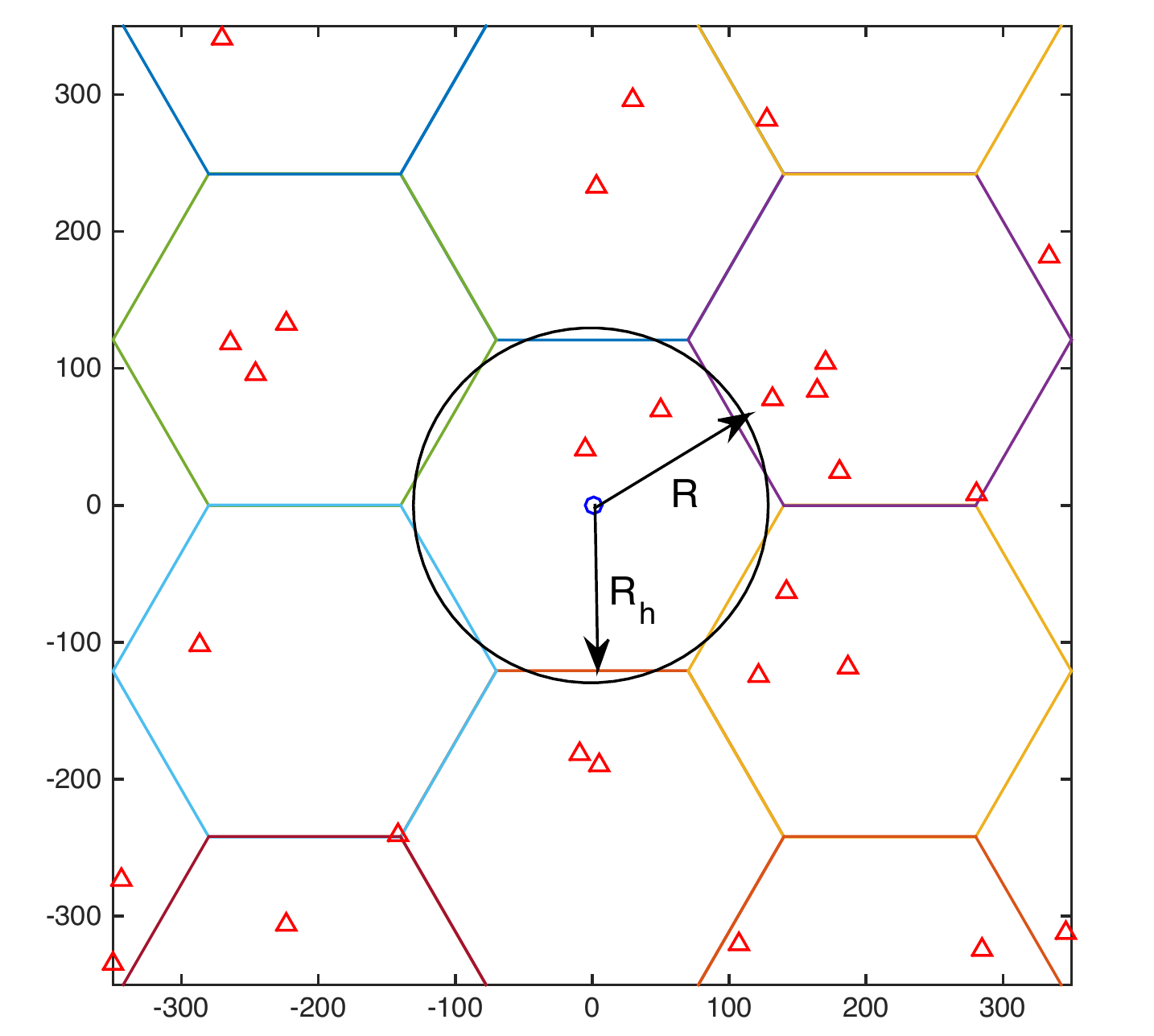}
	\caption{ A snapshot of the network topology of the considered cluster-centric SCN. A hexagonal grid defines the clusters, wherein SBSs (red triangles) are distributed according to a homogeneous PPP. A cluster of interest is considered for performance analysis with cluster center at the origin.}
	\label{network}
\end{figure}

\subsection{Small Cell Clustering}
We consider a cache-enabled SCN where SBSs are distributed according to a homogeneous Poisson point process (PPP) $\Phi_b= \{b_i \in \mathbb{R}^2, \forall i\in \mathbb{N}^{+}\}$ with intensity $\lambda_b$. Nearby SBSs are grouped into disjoint clusters modeled using a hexagonal grid with inter-cluster center distance equal to $2R_{\text{h}}$ \cite{hexagonal}, as shown in Fig. \ref{network}. SBSs belonging to the same cluster can cooperate in order to serve users inside the cluster. The total cache (storage) capacity in a cluster is considered as an entity and cache placement decisions are performed at the central controllers (CCs). The CCs are located at the center of each cluster with positions denoted by $\mathcal{H}=\{y_j \in \mathbb{R}^2, \forall j\in \mathbb{N}^{+}\}$. The area of each cluster is given by $\mathcal{A}=2\sqrt{3}R_{\text{h}}^2$. 
For a random hexagonal cluster, the probability mass function (pmf) of the number $n$ of SBSs in a cluster, which follows a Poisson distribution with mean $\lambda_b\mathcal{A}$, is given by 
\begin{equation}
\mathbb{P}(n=K)=e^{-2\sqrt{3}\lambda_b R_{\text{h}}^2}  \frac{\left(2\sqrt{3}\lambda_b R_{\text{h}}^2\right)^{K}}{K !}.
\label{pmf}
\end{equation}
In clusters with no SBS inside, i.e., $n=0$, users connect to the nearest SBS to download the requested content. For simplicity, we do not consider the case of empty clusters.\footnote{In this work, our main interest is the cache content placement in a cluster-centric SCN. When the cluster is empty, there is no cache placement to perform. Hence, this case can be ignored in our analysis.}

Conditioning on having $K$ SBSs in the cluster of interest with cluster center $y_0$ at the origin, the in-cluster SBS distribution follows a binomial point process (BPP), which consists of $K$ uniformly and independently distributed SBSs in the hexagonal cluster. The distance distribution between randomly distributed nodes and the cell center for hexagonal cell can be found in \cite{hexagon_pdf}. For analytical convenience, we approximate the cluster area to a circle with the same area, i.e., with radius $R=R_{\text{h}}\sqrt{\frac{2 \sqrt{3}}{\pi}}$, as shown in Fig. \ref{network}. The set of cooperative SBSs inside the cluster of interest is thus defined as $\mathcal{C}=\{b_i \in \Phi_b \cap \mathcal{B}(y_0, R)\}$, where $\mathcal{B}(y_0, R)$ denotes the ball centered at $y_0$ with radius $R$. This approximation turns out to have negligible impact on the performance of the network under study \cite{hexagon_pdf}. Consider a user located at the origin (cell center), the distances from the cooperative SBSs to the user are denoted by $\mathbf{r}=[r_1, r_2, \ldots, r_K]$. The $K$ cooperative SBSs can be approximately seen as the $K$ closest SBSs to the cluster-center user.

\subsection{Cache Placement Strategies}
We consider a finite content library $\mathcal{F}=\{f_1, \ldots, f_N\}$, where $N$ is the library size and $f_m$ is the $m$-th most popular file with normalized size equal to $1$. Each user makes independent request for a file with probability according to a given popularity pattern, e.g., Zipf distribution, which is a commonly used distribution for video popularity \cite{zipf}. Suppose we have the request probability of each file in $\mathcal{F}$ denoted by $\mathbf{p}=\{p_1, \ldots, p_N\}$.
With Zipf distribution, the request probability of the $m$-th most popular file is given as
\begin{equation}
p_m=\left(m^{\gamma} \sum\limits_{n=1}^{N}n^{-\gamma}\right)^{-1}, 
\label{zipf_probability}
\end{equation}
where $\gamma$ is the shape parameter, denoting the popularity skewness. 

Due to finite caching capacity, each SBS can store up to $M$ files. In a cluster with $K$ cooperative SBSs, the total available storage capacity is $KM$. Each video file is divided into $K$ partitions and every partition contains the same number of segments. In our cluster-centric SCN model, we consider a combined ``most popular content" (MPC) and ``largest content diversity" (LCD) caching strategy with partition-based caching \cite{partition_based} to distribute partitions of content to the SBSs in the same cluster.\footnote{Partition-based caching, as an example of segment-based caching, gives potential opportunity for faster content delivery brought by parallel transmission.}
Specifically, a proportion $\rho$ of cache space in each SBS is used for caching the most popular content, and the rest $1-\rho$ proportion is reserved for disjointly placing different partitions of the less popular files in different SBSs to increase the content diversity. Hence, files $f_m$ with popularity order $1\leq m\leq \lfloor\rho M \rfloor $ are cached in every SBS inside the cluster (i.e., MPC-based caching). For files $f_m$ with $\lfloor\rho M \rfloor < m\leq \lfloor \rho M\rfloor+K(M-\lfloor \rho M\rfloor) $, every SBS has one different partition of each file (i.e., LCD-based caching). For $m>\lfloor \rho M\rfloor+K(M-\lfloor \rho M\rfloor)$, the files are not cached. In total $\lfloor \rho M\rfloor+K(M-\lfloor \rho M\rfloor)$ different files can be cached inside a cluster.

\begin{figure}
	\centering
	\subfigure[Joint transmission]{
		\includegraphics[scale=0.3]{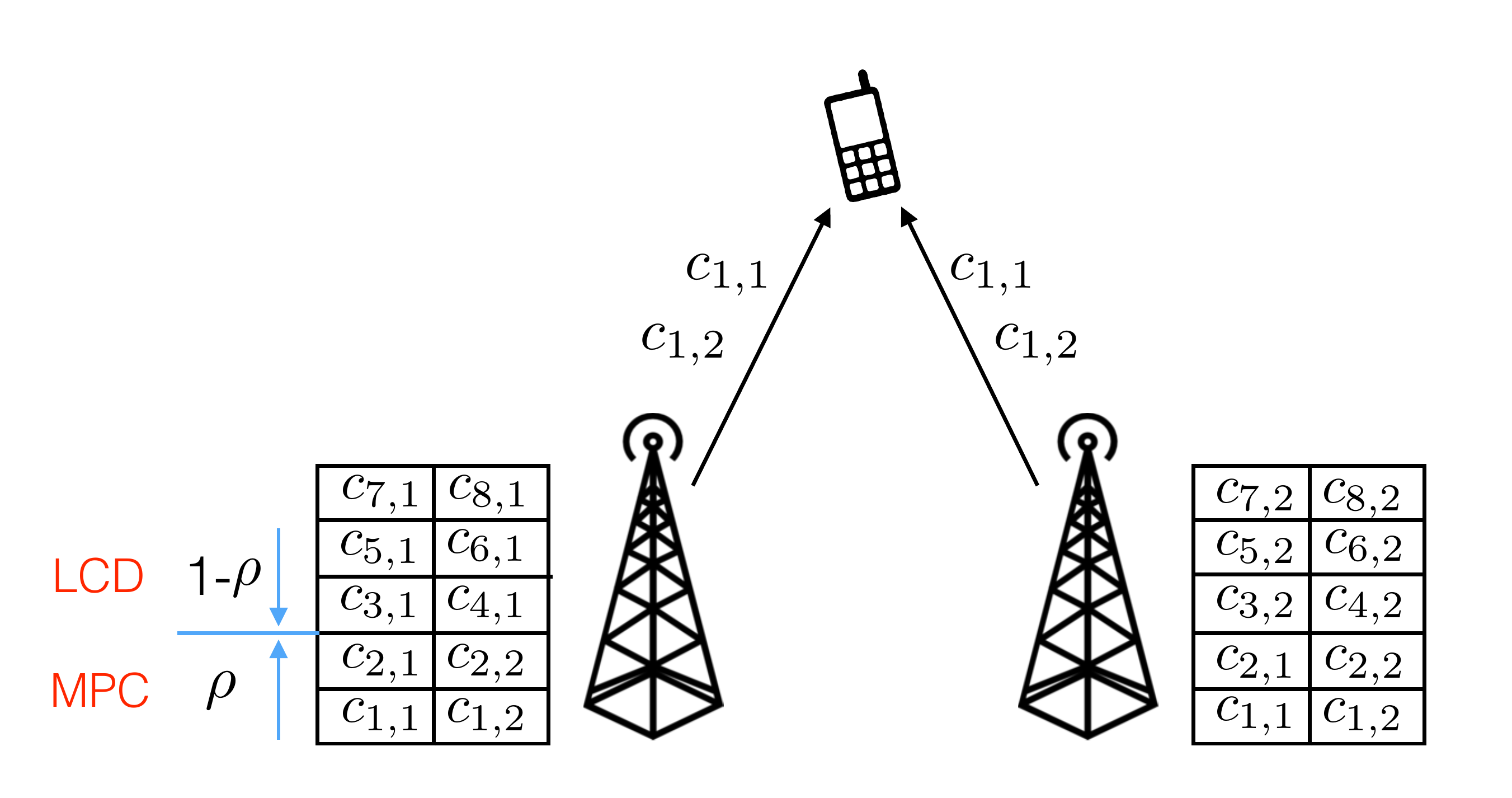}
		\label{mpc}
	}
	\subfigure[Parallel transmission]{
		\includegraphics[scale=0.3]{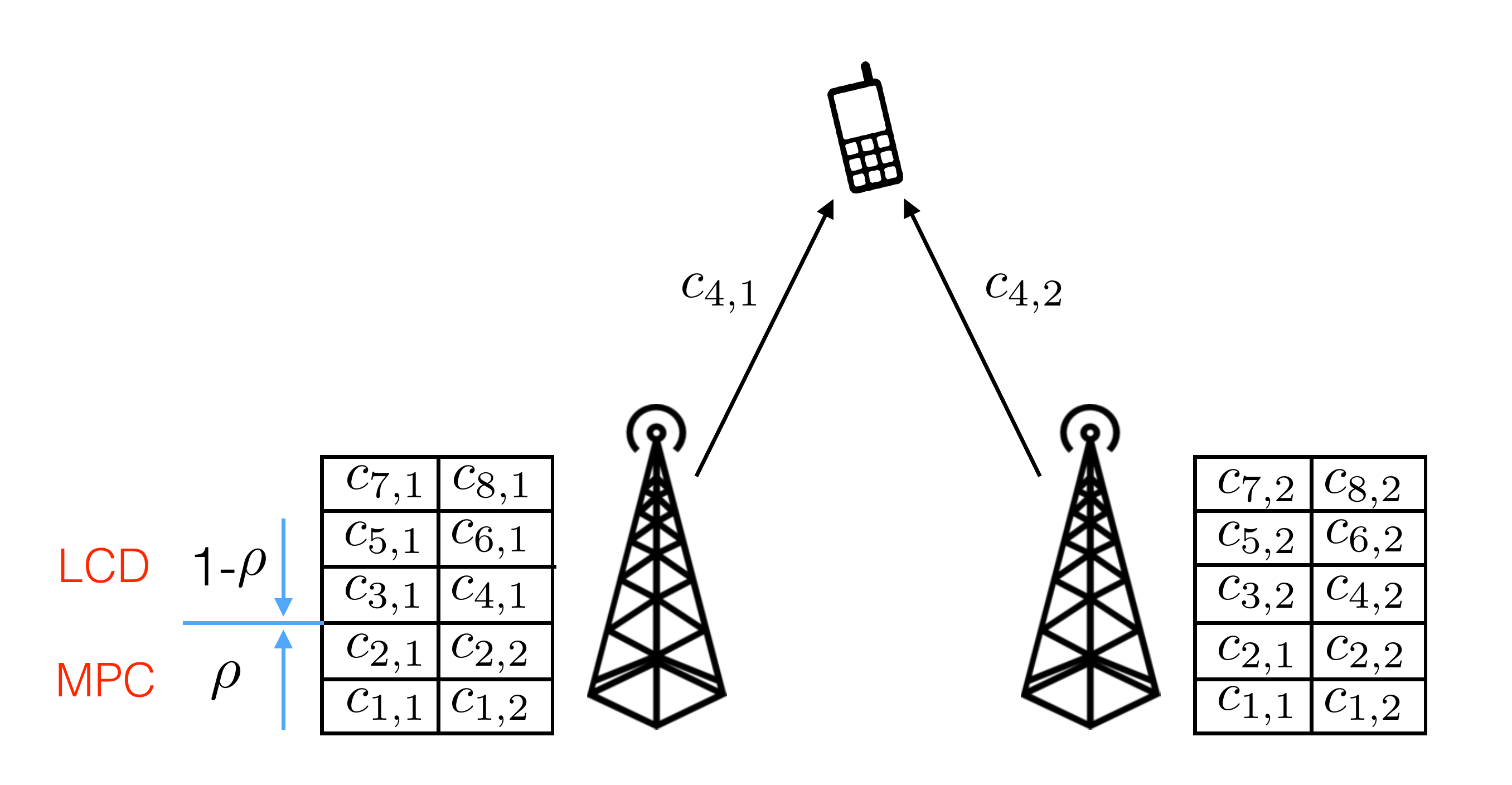}
		\label{lcd}
	}
	\caption{Illustration of combined MPC and LCD caching strategy when $K=2$. Cooperative SBSs use joint transmission/parallel transmission schemes when a user requests for a file falling into the MPC or the LCD range. Here, $c_{i,j}$ denotes the $j$-the partition of the $i$-th file.}
\end{figure}

For a random request within the content library $\mathcal{F}$, the {\emph{cache hit}} probability, i.e., the probability to find the requested file stored in the local cache, is given by
\begin{align}
P_{\text{hit}}(\rho)
&=\sum\limits_{m=1}^{\lfloor \rho M\rfloor+K(M-\lfloor \rho M\rfloor)} p_m \nonumber \\
&=\sum\limits_{m=1}^{\lfloor \rho M\rfloor+K(M-\lfloor \rho M\rfloor)}  \left(m^{\gamma} \sum\limits_{n=1}^{N}n^{-\gamma}\right)^{-1}. 
\label{proba_hit}
\end{align} 
The cache hit probability is a monotonically decreasing function of $\rho$. To increase content diversity, more cache (storage) space should be reserved for the LCD-based caching.

\subsection{Transmission Schemes}
\label{section_transmission}
When a random user inside a cluster requests for a file in $\mathcal{F}$, since the availability of the requested file at in-cluster SBSs differs with the popularity order of the file, different transmission schemes, i.e., namely joint transmission (JT) and parallel transmission (PT), are used for delivering the files in the MPC and LCD ranges, respectively, as described below. 

\subsubsection{Joint Transmission}
If the requested file $f_m$ is in the MPC range, i.e., the popularity order is between $1\leq m\leq \lfloor\rho M \rfloor$, $K$ SBSs in the cluster have the same entire file. Hence, the requested file is jointly transmitted to the user as a means to enhance the content delivery reliability, i.e., increase the received SINR, as shown in Fig. \ref{mpc}. We denote this case as JT cooperation scheme. 

\subsubsection{Parallel Transmission}
If the requested file $f_m$ is in the LCD range, i.e., the popularity order is between $\lfloor\rho M \rfloor < m\leq \lfloor \rho M\rfloor+K(M-\lfloor \rho M\rfloor) $, cooperating SBSs inside the same cluster have disjoint partitions of the requested file. The different partitions need to be transmitted to the user at the same time by parallel (multiple) streams, one from each cooperating SBS, as shown in Fig. \ref{lcd}. We denote this case as PT cooperation scheme.
There are two ways of frequency allocation: i) PT with orthogonal spectrum assignment (PT-OS) case, each SBS uses $\frac{1}{K}$ of the overall available spectrum to transmit the stored partition of the requested file to the user; and ii) PT with successive decoding based spectrum sharing (PT-SS) case, 
$K$ SBSs concurrently transmit $K$ streams containing different partitions of the requested file to the user using the same available spectrum resources. 
In the PT-SS case, at the receiver, successive decoding with (multi-stream) interference cancellation (SIC) is used to decode the signal according to the received signal power order \cite{SIC, zhang2014}. More explicitly, the strongest signal is decoded first and extracted from the received signal, then proceed to the next decoding layer for the next strongest signal, and so on.

\subsubsection{Transmission for Cache Miss Case}
If the requested file is not cached in local cluster, a {\emph{cache miss}} event occurs. In this case, all BSs fetch the requested content from the core network through backhaul links and jointly transmit the content to the user in order to reduce the delivery latency. The power consumption consists of the required power for fetching content from the core network to the cooperative SBSs and the transmission power for delivering content from the cooperative SBSs to the user. 
The backhauling process increases not only end-to-end delivery delay but also energy consumption, compared to the case of serving user requests using the local caches \cite{backhaul_2, ee_hetnets}. By considering those impacts, the energy efficiency of this case is investigated in Section \ref{ee_optimization}.
%

\section{Successful Content Delivery Probability (SCDP) Analysis}
\label{SCDP}
In this section, we study a key metric for the performance evaluation of cluster-centric caching with SBS cooperation, namely the successful content delivery probability (SCDP). We give analytical results on the SCDPs of JT, PT-SS and PT-OS cases for a user located at the cluster center. Note that taking the cluster-center user as a reference is mainly done for analytical tractability, but it can be seen as an upper bound on the SCDP for randomly located users inside the cluster of interest. 

\subsection{SCDP Definition}
Assuming that each file contains $S$ bits, the successful delivery of a file is defined by the event that $S$ bits are successfully delivered using bandwidth $W$ and time $T$. Note that in the JT and PT cases, the numbers of information bits delivered from each SBS are different.
In the JT case, each SBS sends $S$ bits to the user using the same bandwidth. Hence, at the receiver side, the received signals from $K$ SBSs are superimposed and thus considered as a single stream. The SCDP is defined as a function of the received signal-to-interference-plus-noise ratio (SINR), given as 
\begin{equation}
p_{\text{d},K}^{\text{JT}}=\mathbb{P}\left[W T\log_2(1+\text{SINR})> S \;\middle|\; K\right].
\end{equation}

In the PT-SS case, each SBS sends $\frac{S}{K}$ bits to the user employing SIC by sharing the same $W$ bandwidth.  The decodability of the received streams depends on the SINR of each stream and rate requirement. 
Decoding $K$ streams using SIC is theoretically feasible if all $K$ streams achieves higher rate than the target rate for successful transmission\cite{SIC}. Hence, we have
\begin{equation}
p_{\text{d},K}^{\text{PT-S}}=\mathbb{P}\left[ \bigcap\limits_{i\in \{1,\ldots, K\}} W T\log_2(1+\text{SINR}_i)> \frac{S}{K} \;\middle|\; K\right],
\end{equation}
where $\text{SINR}_i$ is the received SINR of the stream containing the $i$-th partition of the requested file. 

In the PT-OS case, each BS sends $\frac{S}{K}$ bits by using $\frac{W}{K}$ bandwidth each. The delivery rate is bounded by the stream with the lowest achievable rate, so the SCDP is defined as
\begin{equation}
p_{\text{d},K}^{\text{PT-O}}=\mathbb{P}\left[\frac{W}{K} T\log_2(1+\min\limits_{i\in[1,\ldots, K]}\{\text{SINR}_i\})> \frac{S}{K} \;\middle|\; K \right].
\end{equation}

We denote $R_d=\frac{S}{T}$ (bit/s) as the target rate for successful content delivery. In terms of SINR requirement, the SCDP can be rewritten as 
\begin{eqnarray}
&& p_{\text{d},K}^{\text{JT}}=\mathbb{P}\left[\text{SINR}> 2^{\frac{R_d}{W}}-1 \;\middle|\; K\right] \label{def_scdp_jt}\\
&& p_{\text{d},K}^{\text{PT-S}}=\mathbb{P}\left[\bigcap\limits_{i\in \{1,\ldots, K\}}\text{SINR}_i> 2^{\frac{R_d}{KW}}-1\;\middle|\; K\right]  \label{def_scdp_ptss} \\
&& p_{\text{d},K}^{\text{PT-O}}=\mathbb{P}\left[\min\limits_{i\in[1,\ldots, K]}\{\text{SINR}_i\}> 2^{\frac{R_d}{W}}-1\;\middle|\; K\right ],  \label{def_scdp_ptos}
\end{eqnarray}
for $i=1,\ldots, K$.

\subsection{SCDP of MPC-JT strategy} 
\label{JT}
For the cluster-center user located at $y_0=\left(0, 0\right) $, when it requests for file $f_m$ with $1\leq m\leq \lfloor\rho M \rfloor$, which is in the MPC range, coordinated joint transmission is used to combine coherently the received signals from cooperating SBSs. Hence, over each symbol duration time, the cooperating SBSs transmit the same symbol $s$. Assuming equal transmit power $P_t$ for every SBS and a standard distance-dependent power law pathloss attenuation, i.e., $r^{-\alpha}$, where $\alpha>2$ is the pathloss exponent, the channel output at the user is
\begin{equation}
y=\sum_{b_i\in \mathcal{C}} \sqrt{P_t} {r_i}^{-\frac{\alpha}{2}} h_i s +\sum_{b_j\in \Phi_b \backslash \{\mathcal{C}\} } \sqrt{P_t} {r_j}^{-\frac{\alpha}{2}} h_j s_j+n,
\end{equation} 
where $h_l$ denotes the small-scale Rayleigh fading from the $l$-th SBS to the user, which follows $h_l \sim \mathcal{CN}(0, 1)$; $r_l$ denotes the distance from the $l$-th SBS to the user; $s_l$ denotes the transmitted symbol of the $l$-th SBS; and $n$ denotes the background thermal noise.

Considering an interference-limited network and neglecting the background thermal noise, the signal-to-interference ratio (SIR) of received signal is given by
\begin{equation}
\text{SIR}_{\text{JT}}=\frac{ \left|\sum\limits_{b_i\in \mathcal{C}}h_i {r_i}^{-\frac{\alpha}{2}}\right|^2}{\sum\limits_{b_j\in \Phi_b \backslash \{\mathcal{C}\}} |h_j|^2 {r_j}^{-\alpha}}.
\label{sir_jt}
\end{equation} 
Using  \eqref{def_scdp_jt} and \eqref{sir_jt}, we can obtain the SCDP of JT case as follows.

\begin{lemma}
	\label{lemma1}
	For the cluster-center user with target SIR $\theta_1=2^{\frac{R_b}{W}}-1$,
	the SCDP of JT case with $K$ cooperating SBSs is given by
	\begin{equation}
	\begin{split}
	&p_{\textnormal{d},K}^{\textnormal{JT}}(\theta_1)  \\
	& \simeq \int_{0}^{R}\!\!\cdots\int_{0}^{R} \mathcal{L}_{I|R} \left(\frac{\theta_1}{\sum_{i=1}^{K}x_i^{-\alpha}}\right) \prod\limits_{i=1}^{K} \frac{2 x_i}{R^2} \textnormal{d}x_1 \cdots \textnormal{d}x_K,
	\end{split}
	\label{Psuc_JT_expression}
	\end{equation} 
	where $\mathcal{L}_{I|x} (s)$ is the Laplace transform of the interference coming from SBSs located outside of $\mathcal{B}(0,x)$, given by
	\begin{equation}\label{eq:Laplace}
	\mathcal{L}_{I|x}(s)=\exp\left(-\pi \lambda_b s^{\frac{2}{\alpha}} \int_{\frac{x^2}{s^{2/\alpha}}}^{\infty} \frac{1}{1+w^{\frac{2}{\alpha}}} \textnormal{d}w\right).
	\end{equation}
\end{lemma}
\begin{IEEEproof}
	\textnormal{See Appendix \ref{appen1}.}
\end{IEEEproof}

\subsection{SCDP of LCD-PT strategy} 
When the cluster-center user requests for file $f_m$ with $\lfloor\rho M \rfloor < m\leq \lfloor \rho M\rfloor+K(M-\lfloor \rho M\rfloor) $, which is in the LCD range, parallel streams containing different partitions of the requested file are simultaneously sent to the user. Considering different spectrum usages, we study SCDPs for PT-SS and PT-OS cases separately in this section.
\subsubsection{PT-SS}
\label{PT}
In the PT-SS case, over each symbol duration time, $K$ SBSs transmit $K$ different symbols (one symbol in each partition) $[s_1, s_2, \ldots, s_K]$ to the user at the origin. If all SBSs use the same transmit power $P_t$ as in the JT case, the channel output at the receiver is
\begin{equation}
y=\sum\limits_{b_j\in \mathcal{C}} \sqrt{P_t} {r_i}^{-\frac{\alpha}{2}} h_i s_{i} +\sum_{b_j\in \Phi_b \backslash \{\mathcal{C}\} } \sqrt{P_t} {r_j}^{-\frac{\alpha}{2}} h_j s_j+n.
\end{equation} 
In order to decode multiple streams simultaneously, we use SIC with respect to a certain order of received signal.
The detailed analysis of SIC based on power ordering statistics is out of the scope of this paper and has been studied in \cite{zhang2014}. For the ease of analysis, we consider here the case where the user decodes different information streams based on the distance order~\cite{successive_ic}. After approximating the cluster area by the circle $\mathcal{B}(0, R)$, the decoding order will be from the nearest SBS to the $K$-th nearest SBS to the cluster-center user.
We define $\widetilde{\mathbf{r}}=[\widetilde{r}_1, \ldots, \widetilde{r}_K]$ the distance vector with increasing distance order, where $\widetilde{r}_k, k\in[1,K]$ is the distance from the $k$-th nearest SBS to the cluster-center user. 

When decoding the information from the $k$-th nearest SBS, all signals coming from closer SBSs $\{b_1,\ldots, b_{k-1}\}$ need to be successfully decoded and canceled.
In this case, the interference comes from $K-k$ remaining SBSs inside the cluster and PPP distributed SBSs outside the cluster. Due to the conditioned number $K$, the interference distribution is different from the case with PPP-distributed SBSs.
For the tractability analysis, we assume that at the $k$-th decoding step with $k\in[1, K-1]$, the distribution of interfering SBSs outside $\mathcal{B}(0, \widetilde{r}_k)$ still follows a homogeneous PPP. The SIR of the $k$-th stream with SIC is thus given as
\begin{equation}
\text{SIR}_{k}\simeq\frac{\left|h_k\right|^2 {\widetilde{r}_k}^{-\alpha}}{\sum\limits_{b_j\in \Phi_b \backslash \mathcal{B}(0, \widetilde{r}_k)} |h_j|^2 {r_j}^{-\alpha}}.
\label{PTSS_approximation}
\end{equation} 

At the last decoding step, all in-cluster interfering signals are canceled. The remaining interference comes from out-of-cluster SBSs with minimum distance $R$ to the user. Hence, for the last decoded stream, we have 
\begin{equation}
\text{SIR}_{K}\simeq \frac{\left|h_K\right|^2 {\widetilde{r}_K}^{-\alpha}}{\sum\limits_{b_j\in \Phi_b \backslash \mathcal{B}(0, R)} |h_j|^2 {r_j}^{-\alpha}}.
\label{PTSS_approximation2}
\end{equation} 
Using \eqref{def_scdp_ptss}, \eqref{PTSS_approximation}, and \eqref{PTSS_approximation2}, we now obtain the SCDP of PT-SS case as follows.

\begin{lemma}
	\label{lemma2}
	For the cluster-center user with target SIR $\theta_2=2^{\frac{R_d}{KW}}-1$,
	the SCDP of PT-SS case with $K$ cooperating SBSs is given by
	\begin{align}
	\begin{split}
	p_{\textnormal{d},K}^{\textnormal{PT-S}}(\theta_2)
	\simeq  
	& \int\limits_{0<x_1<\cdots<x_K<R} 
	\frac{2 K\cdot x_K}{R^2} \mathcal{L}_{I|R}\left(\theta_2 x_K^\alpha\right)\\ 
	& \quad\times \prod\limits_{k=1}^{K-1} \frac{2 k\cdot x_k}{R^2} \mathcal{L}_{I|x_k}\left(\theta_2 x_k^\alpha\right)
	\textnormal{d}x_1 \cdots \textnormal{d}x_K ,
	\end{split}
	\label{Psuc_PT_expression}
	\end{align}
	where $\mathcal{L}_{I|x}(s)$ is defined in \eqref{eq:Laplace}. 
\end{lemma}
\begin{IEEEproof}
	\textnormal{See Appendix \ref{appen2}.}
\end{IEEEproof}

\subsubsection{PT-OS}
In the PT-OS case, different SBSs transmit different partitions of the requested content through orthogonal frequency bandwidth. For the information stream transmitted from the $i$-th SBS, we have the channel output as
\begin{equation}
y_{i}=\sqrt{P_t} {r_i}^{-\frac{\alpha}{2}} h_{i} s_{i} +\sum_{b_j\in \Phi_b \backslash \{\mathcal{C}\} } \sqrt{P_t} {r_j}^{-\frac{\alpha}{2}} h_j s_j+n.
\end{equation} 
Then the received SIR of the $i$-th stream is 
\begin{equation}
	\text{SIR}_{i}=\frac{\left|h_i\right|^2 {r_i}^{-\alpha}}{\sum\limits_{b_j\in \Phi_b \backslash \mathcal{C}} |h_j|^2 {r_j}^{-\alpha}}.
	\label{sir_ptos}
\end{equation} 
Using \eqref{def_scdp_ptos} and \eqref{sir_ptos}, we obtain the SCDP of PT-OS case as follows.

\begin{lemma}
	\label{lemma3}
	For the cluster-center user with target SIR $\theta_1=2^{\frac{R_d}{W}}-1$,
	the SCDP of PT-OS case with $K$ cooperating SBSs is given by 
	\begin{equation}
	p_{\textnormal{d},K}^{\textnormal{PT-O}} (\theta_1)  
	\simeq  
	 \int_{0}^{R}\!\!\cdots\int_{0}^{R}  \prod\limits_{i=1}^{K} \frac{2 x_i}{R^2}.  \mathcal{L}_{I|R}\left(\theta_1 x_i^{\alpha}\right)\textnormal{d}x_1 \cdots \textnormal{d}x_K,
	\label{Psuc_pt_os}
	\end{equation}  	
	where $\mathcal{L}_{I|x}(s)$ is defined in \eqref{eq:Laplace}. 
\end{lemma}
\begin{IEEEproof}
	\textnormal{See Appendix \ref{appen3}.}
\end{IEEEproof}

\section{Optimization Design of Cache Utilization Strategy}
\label{optimal caching}
In this section, we first show the inherent tradeoff between transmission diversity and content diversity based on our analysis in Section \ref{SCDP}. We then define two optimization problems in order to provide the optimal cache space assignment for the proposed combined caching strategy.

\subsection{Transmission Diversity vs. Content Diversity}	
\label{tradeoff}
\begin{figure}
	\centering
	\includegraphics[scale=0.45]{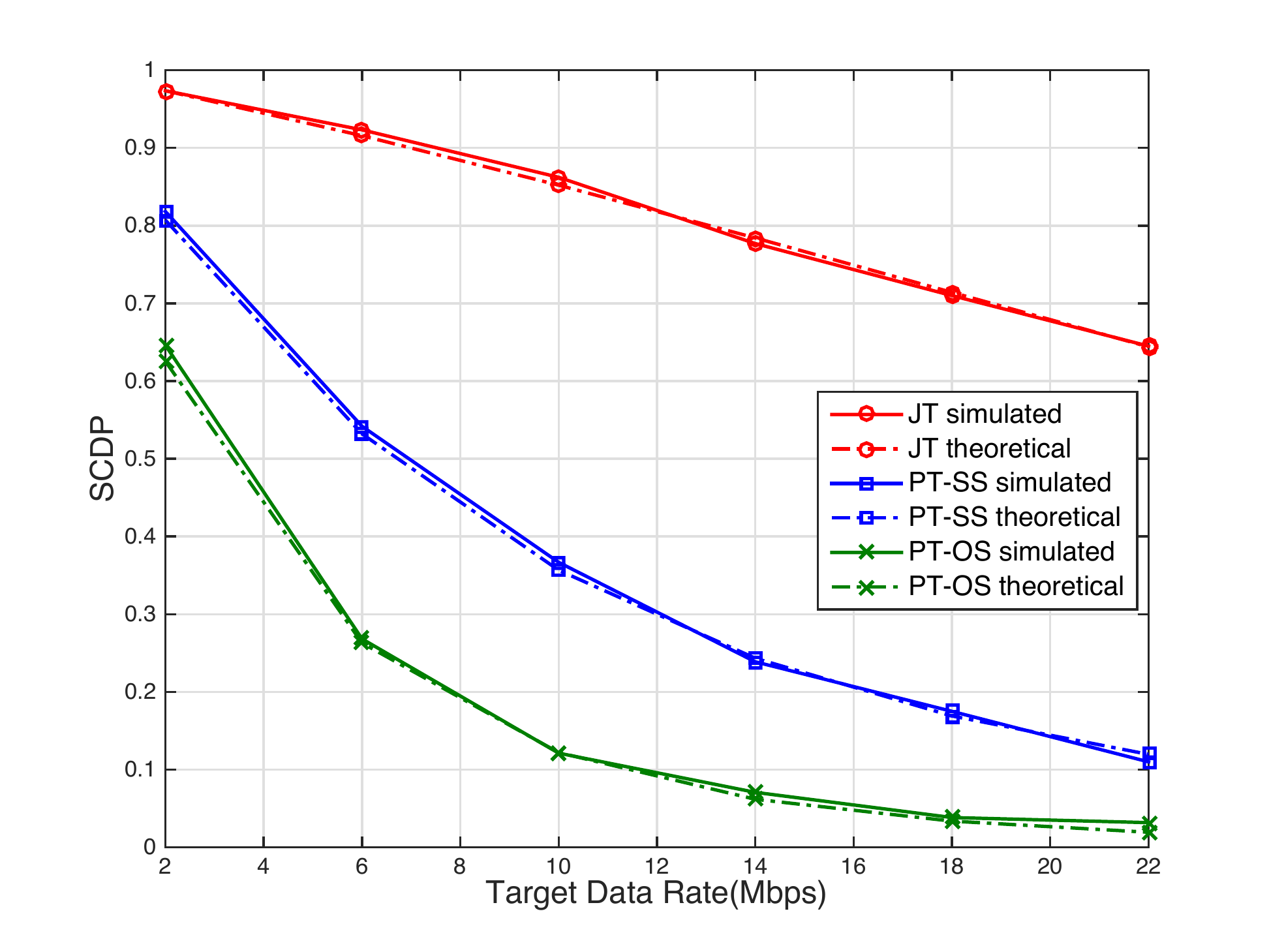}
	\caption{Theoretical and simulation results of SCDP of JT, PT-SS and PT-OS transmission schemes.}
	\label{Psuc_JT_PT}
\end{figure}

\begin{figure}
	\centering
	\includegraphics[scale=0.45]{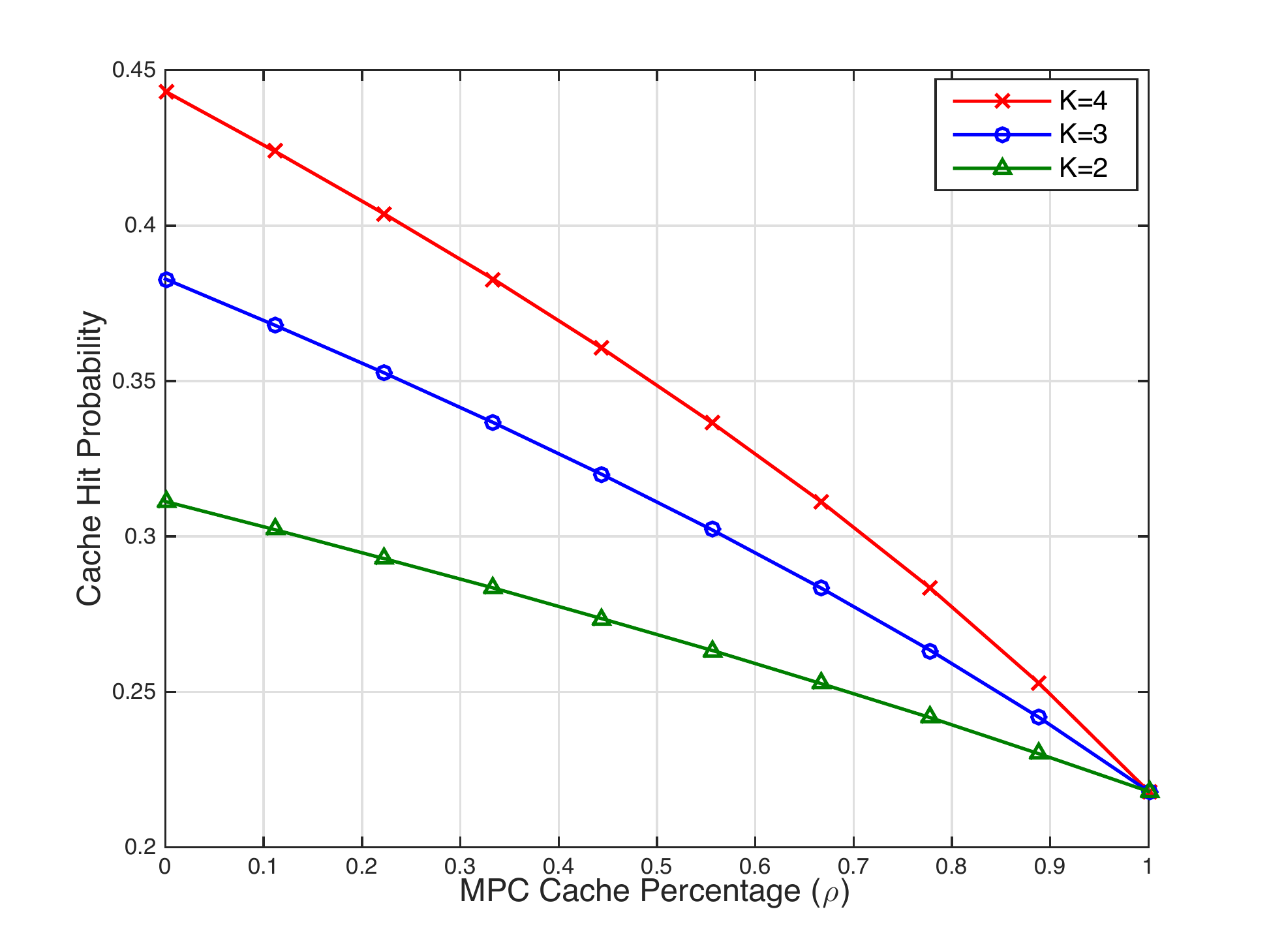}
	\caption{Cache hit probability vs. MPC cache percentage ($\rho$) for $K=2, 3, 4$. $\gamma=0.5$.}
	\label{Cache_hit}
\end{figure}

In Fig. \ref{Psuc_JT_PT}, we plot both numerical and simulation results of SCDP of the three transmission schemes discussed in Section~\ref{SCDP} as a function of the target rate. The theoretical values are obtained from \eqref{Psuc_JT_expression}, \eqref{Psuc_PT_expression} and \eqref{Psuc_pt_os} for the JT, PT-SS, and PT-OS cases, respectively. For the simulation results, the values of used parameters are given in Table~\ref{system_params} of Section~\ref{sec:simulation}. The number of cooperative SBSs is chosen as $K=3$, which is close to the average number of SBSs per cluster according to our network settings. 
Fig. \ref{Cache_hit} shows the cache hit probability given in \eqref{proba_hit} as a function of the percentage of cache space assigned for MPC caching strategy in each SBS, $\rho$. 

From~Fig.~\ref{Psuc_JT_PT}, we first see that simulation results of SCDP match well with numerical results.
We also observe that JT always achieves higher SCDP than PT cases, evincing the benefit of MPC caching and JT transmission scheme in terms of higher transmission reliability.
In addition, PT-SS always has higher SCDP than PT-OS, because spectrum sharing with SIC gives better reuse of communication resources for the parallel transmission. Therefore, in the following, we only consider PT-SS as the transmission scheme when the requested content falls in LCD range. 
Hence, when we refer to PT transmission scheme, it means PT-SS scheme.  

From Fig.~\ref{Cache_hit}, we may observe that for lower $\rho$ we have better cache hit ratio due to higher content diversity, achieved by assigning more space for LCD caching.
From those two figures, we see that higher $\rho$ increases the chance for joint transmission, which helps to improve the transmission reliability. With lower $\rho$, more different files will be cached in the cluster, thus offering higher cache hit probability.  
In other words, there is a tradeoff between transmission diversity and content diversity. The cluster-centric cache utilization design should be able to leverage both diversity gains and adapt to the network environment and requirements. For instance,
when the transmission rate requirement is high, caching the same most popular files in every SBS is preferable. Alternatively, increasing content diversity brings more opportunities to handle local requests by the cache.

\subsection{Optimal Design for Cache Service Performance }
\label{cache_service_optimization}
Since the MPC cache percentage $\rho$ in each SBS affects both local content diversity and transmission reliability, we seek here the optimal $\rho$ that maximizes the percentage of requests successfully served by local caches, namely the cache service probability. A user request can be successfully served by local caches only when: 1) the requested file is cached inside the cluster, 2) the content delivery from the cooperative SBSs to the user is successful. We then define the cache service probability as follows.

\begin{definition}
	In the cluster-centric SCN with proposed combined caching strategy, the average cache service probability is given as
	\begin{equation}
	p_{\textnormal{sv}}=\sum\limits_{K=1}^{\infty} \mathbb{P}(n=K) f(\rho| K),
	\label{average_psv}
	\end{equation}
	where $f(\rho| K)$ is the cache service probability conditioning on having $K$ SBSs inside the cluster, given by 
	\begin{equation}
	f(\rho|K) = p_{\textnormal{CH, M}}(\rho)\, p_{\textnormal{d},K}^{\textnormal{JT}} (\theta_1) + p_{\textnormal{CH, L}}(\rho)\,  p_{\textnormal{d},K}^{\textnormal{PT-S}} (\theta_2).
	\label{objectif_function}
	\end{equation}
$p_{\textnormal{d},K}^{\textnormal{JT}} (\theta) $ and $ p_{\textnormal{d},K}^{\textnormal{PT-S}} (\theta)$ are given in \eqref{Psuc_JT_expression} and \eqref{Psuc_PT_expression}, respectively. 
Here, $p_{\textnormal{CH, M}}(\rho)$ and $p_{\textnormal{CH, L}}(\rho)$ are the probabilities to have the requested file cached in the MPC and LCD ranges, respectively, given by 
\begin{align}	
	p_{\textnormal{CH, M}}(\rho)& =\sum\limits_{m=1}^{\lfloor \rho M\rfloor } p_{m} \label{Cache_Prob_MPC},\\
	p_{\textnormal{CH, L}}(\rho)& =\sum\limits_{\lfloor \rho M\rfloor+1}^{\lfloor \rho M\rfloor+K(M-\lfloor \rho M\rfloor) } p_{m}, \label{Cache_Prob_LCD}
\end{align}
where $p_m$ is defined in \eqref{zipf_probability}.
\end{definition}

Since each cluster performs cooperative caching independent of other clusters, for a random cluster with $K$ SBSs, the objective is to maximize its in-cluster cache service probability, that is, to find $\rho$ which maximizes $f(\rho|K) $.

When $\gamma<1$ and $M\ll N$, since $p_m \propto 1/ m^{\gamma}$, we have \cite{robert}
\begin{equation}
g(L)=\sum\limits_{m=1}^{L} p_{m}\approx (L/N)^{1-\gamma} .
\label{approximation}
\end{equation}
Using this approximation, $p_{\textnormal{CH, M}}(\rho)$ and $p_{\textnormal{CH, L}}(\rho)$ can be approximated by two continuous functions of $\rho$, given by
\begin{align}	
p_{\textnormal{CH, M}}(\rho)& \simeq \left(\frac{M}{N}\right)^{1-\gamma} \rho^{1-\gamma} = \widetilde{p}_{\textnormal{CH, M}}(\rho), \label{Cache_Prob_MPC_approxi}\\
\begin{split}
p_{\textnormal{CH, L}}(\rho)& \simeq \left(\frac{M}{N}\right)^{1-\gamma} \left\{\left[\rho(1-K)+K\right]^{1-\gamma}-\rho^{1-\gamma} \right\}\\
&= \widetilde{p}_{\textnormal{CH, M}}(\rho).
\end{split}
 \label{Cache_Prob_LCD_approxi}
\end{align}
Then, the cache service probability in \eqref{objectif_function} is simplified as
\begin{equation}
\begin{split}
f(\rho|K)&\simeq \left(\frac{M}{N}\right)^{1-\gamma} \rho^{1-\gamma} p_{\textnormal{d},K}^{\text{JT}}(\theta_1)  \\
&+ \left(\frac{M}{N}\right)^{1-\gamma} \left[\left(\rho(1-K)+K\right)^{1-\gamma}-\rho^{1-\gamma} \right]p_{\text{d},K}^{\text{PT-S} }(\theta_2).
\end{split}
\label{approximate_function}
\end{equation}
Using \eqref{approximate_function}, we can obtain the optimal $\rho$ as follows.

\begin{lemma}
	\label{lemma4}
	In a cluster-centric SCN with proposed combined caching strategy, knowing that there are $K$ cooperative SBSs in the cluster, the optimal percentage of cache space assigned for MPC caching is given as
	\begin{align}
	\rho^{*}&= \argmax_{\rho \in[0,1] } f(\rho|K) \nonumber\\
	&\simeq \min \left\{K\left[\left(\frac{K-1}{\frac{p_{\textnormal{d},K}^{\textnormal{JT}} (\theta_1) }{p_{\textnormal{d},K}^{\textnormal{PT-S}} (\theta_2) }-1}\right)^{1/\gamma}+K-1\right]^{-1} ,1\right\},
	\label{optimal_rho}
	\end{align}
	where $p_{\textnormal{d},K}^{\textnormal{JT}} (\theta_1)$ and $p_{\textnormal{d},K}^{\textnormal{PT-S}} (\theta_2)$ are given in \eqref{Psuc_JT_expression} and \eqref{Psuc_PT_expression}, respectively.
\end{lemma}
\begin{IEEEproof}
	\textnormal{See Appendix \ref{appen4}.}
\end{IEEEproof}

\begin{remark}
From \eqref{optimal_rho}, we can see that the ratio $\frac{p_{\textnormal{d},K}^{\textnormal{JT}} (\theta_1) }{p_{\textnormal{d},K}^{\textnormal{PT-S}}(\theta_2)}$ is critical for the optimal cache assignment. When the transmission reliability of JT scheme is much higher than that of PT scheme, i.e., $p_{\textnormal{d},K}^{\textnormal{JT}} (\theta_1)\gg p_{\textnormal{d},K}^{\textnormal{PT-S}} (\theta_2)$, we have $\rho^{*} \simeq1$, meaning that most of the cache space would be used to store the most popular contents. When $\frac{p_{\textnormal{d},K}^{\textnormal{JT}} (\theta_1) }{p_{\textnormal{d},K}^{\textnormal{PT-S}}(\theta_2)}\simeq 1$, $\rho^{*} \simeq 0$, then increasing the content diversity becomes more beneficial. 
\end{remark}

Inside each cluster, based on its knowledge about the number of in-cluster SBSs and out-of-cluster interfering SBS density, the central controllers will be able to compute the optimal percentage of cache space for MPC caching and assist the cache placement in each cooperative SBS.

\subsection{Optimal Design for Energy Efficiency (EE)}
\label{ee_optimization}
When a user requests for a file, depending on the availability of this file in local caches and the placement strategy, both the delivery rate and power consumption will be different. 
If the requested file is not in local caches, the SBSs serving the user needs to download the file from the core network throughout backhaul.
In that case, energy is consumed at the backhaul and there is additional delay of downloading from the core network to the SBSs.
As a result, the energy consumption and the content delivery rate are determined according to our cache utilization design, more explicitly, they depends on $\rho$ in the combined caching scheme.
In our network model, the energy efficiency can be defined as
the effective delivery rate per unit energy consumption, where the effective delivery rate is the number of successfully delivered bits per second, similar to \cite{energy_efficiency}.

When the requested file is stored in local caches (i.e., cache hit case), the effective delivery rate is defined as
$R_d p_{\text{d},K}^{\text{JT}} (\theta_1)$ and $R_d p_{\text{d},K}^{\text{PT-S}} (\theta_2)$
for the MPC-JT and LCD-PT cases, respectively, where $\theta_1=2^{\frac{R_d}{W}}-1$ and $\theta_2=2^{\frac{R_d}{KW}}-1$ are the corresponding target SIRs.
%
When the requested file is not in local caches (i.e., cache miss case), we need to consider the backhaul delay $T_{\text{bh}} (<T)$ to define the effective delivery rate. 
For delivering the requested file within the time slot $T$,
the maximum transmission time should be $T^{'}=T-T_{\text{bh}}=\beta T$, 
where $\beta=1-\frac{T_{\text{bh}}}{T}$ is the fraction of reduced transmission time due to backhaul delay. As mentioned in Section~\ref{section_transmission}, in the cache miss case, the requested file is downloaded from the core network to every in-cluster SBS and joint transmission will be used to serve the user. 
Hence, the effective delivery rate in this case becomes $R_d p_{\text{d},K}^{\text{JT}} (\theta_3)$ with $\theta_3=2^{\frac{R_d}{\beta W}}-1$.


By taking the aforementioned three cases into account, the average effective date rate can be given as
\begin{equation}\label{Effective_Rate}
\begin{split}
\widetilde{R}_{\textnormal{avg}}= & p_{\textnormal{CH, M}}(\rho) R_d \, p_{\text{d},K}^{\text{JT}} (\theta_1)+ p_{\textnormal{CH, L}}(\rho) R_d \,  p_{\text{d},K}^{\text{PT-S}}(\theta_2) \\
 &+p_{\textnormal{CM}}(\rho) R_d \,  p_{\text{d},K}^{\text{JT}}(\theta_3),
\end{split}
\end{equation}
where $p_{\textnormal{CH, M}}(\rho)$ and $p_{\textnormal{CH, L}}(\rho)$ are defined in \eqref{Cache_Prob_MPC} and \eqref{Cache_Prob_LCD}, respectively,
and $p_{\textnormal{CM}}(\rho)$ is the probability of not having the request file cached inside the cluster (i.e., cache miss probability), given by
\begin{align}	
	p_{\textnormal{CM}} (\rho)=1-\sum\limits_{m=1}^{\lfloor \rho M\rfloor+K(M-\lfloor \rho M\rfloor)} p_{m}.
	\label{Cache_Prob_MISS}
\end{align}


For the cache hit case, the consumed power for content delivery contains only the transmit power of the $K$ SBSs if we ignore other static power consumption for the baseband processing, etc.
For the cache miss case, the requested file is fetched from the core network through backhaul, and then transmitted from the $K$ SBSs to the user. Denote $P_b$ as the wireline backhaul power consumption required to handle a user request at a single SBS \cite{backhaul}. Then, we have the average power consumption to serve a user request inside a cluster of $K$ SBSs as \footnote{Here, we do not consider the static power consumption for the baseband processing, site cooling, etc., since this part of consumed power is the same for MPC, LCD and cache miss cases. Adding the static power in the average power consumption is equivalent to having higher transmit power $P_t$ for each SBS in \eqref{Power_Consumption}.}
\begin{align}
\label{Power_Consumption}
P_{\text{avg}} &=K\left\{ \left[p_{\textnormal{CH, M}}(\rho) +p_{\textnormal{CH, L}}(\rho)\right]  P_t+p_{\textnormal{CM}}(\rho) (P_t + P_b)\right\} \nonumber \\
&=K P_t +K P_b p_{\textnormal{CM}}(\rho),
\end{align}
which is averaged over the three cases. From \eqref{Effective_Rate} and \eqref{Power_Consumption}, we can define the EE as follows.

\begin{definition}
	 In the cluster-centric SCN with proposed combined caching strategy, the average EE is given as
	\begin{eqnarray}
    \eta_{\textnormal{EE}}=\sum\limits_{K=1}^{\infty} \mathbb{P}(n=K) \eta(\rho| K),
	\label{average_ee}
	\end{eqnarray}
	where $\eta(\rho| K)$ is the EE conditioning on having $K$ SBSs inside the cluster, given by
	%
	\begin{equation}
	\begin{split}
	& \eta(\rho| K)=\frac{\widetilde{R}_{\textnormal{avg}}}{P_{\textnormal{avg}}} \\
	& \!= \! \frac{R_d  \!\left[p_{\textnormal{CH,M}} (\rho)p_{\textnormal{d},K}^{\textnormal{JT}}\!(\theta_1)  \!+ \! p_{\textnormal{CH,L}}(\rho) p_{\textnormal{d},K}^{\textnormal{PT-S}}\!(\theta_2)   \!+ \! p_{\textnormal{CM}}(\rho) p_{\textnormal{d},K}^{\textnormal{JT}}\!(\theta_3)\!\right] \!}{K P_t +K P_b p_{\textnormal{CM}}(\rho)}.
	\end{split}
	\label{EE_definition}
	\end{equation} 
Here, $p_{\textnormal{d},K}^{\textnormal{JT}} (\theta) $ and $ p_{\textnormal{d},K}^{\textnormal{PT-S}} (\theta)$ are given in \eqref{Psuc_JT_expression} and \eqref{Psuc_PT_expression}, respectively, and 
$p_{\textnormal{CH,M}}(\rho)$, $p_{\textnormal{CH,L}}(\rho)$ and $p_{\textnormal{CM}}(\rho)$ are given in \eqref{Cache_Prob_MPC}, \eqref{Cache_Prob_LCD}, and $ \eqref{Cache_Prob_MISS}$, respectively.
\end{definition}

Inside a cluster with $K$ cooperative SBSs, the optimal cache utilization strategy that maximizes the EE is given by finding $\rho^{*}= \argmax_{\rho \in[0,1] } \eta(\rho| K)$.
Similarly, with the help of the approximation in \eqref{approximation} for the case when $\gamma<1$ and $M\ll N$, we get
\begin{equation}
p_{\textnormal{CM}}(\rho)\simeq 1-\left(\frac{M}{N}\right)^{1-\gamma}\left[\rho(1-K)+K\right]^{1-\gamma} = \widetilde{p}_{\textnormal{CM}}(\rho) .
\label{Cache_Prob_MISS_approxi}
\end{equation}
Putting \eqref{Cache_Prob_MPC_approxi}, \eqref{Cache_Prob_LCD_approxi} and \eqref{Cache_Prob_MISS_approxi} into \eqref{EE_definition}, we obtain the approximated EE $\widetilde{\eta}(\rho| K)$ as a continuous function of $\rho$, given as
\begin{equation}
\begin{split}
& \widetilde{\eta}(\rho| K) \\
\simeq &\frac{R_d  \!\left[\widetilde{p}_{\textnormal{CH,M}} (\rho)p_{\textnormal{d},K}^{\textnormal{JT}}\!(\theta_1)  \!+ \! \widetilde{p}_{\textnormal{CH,L}}(\rho) p_{\textnormal{d},K}^{\textnormal{PT-S}}\!(\theta_2)   \!+ \! \widetilde{p}_{\textnormal{CM}}(\rho) p_{\textnormal{d},K}^{\textnormal{JT}}\!(\theta_3)\!\right] \!}{K P_t +K P_b \widetilde{p}_{\textnormal{CM}}(\rho)}.
\label{EE_approximate}
\end{split}
\end{equation}

Due to the above involved expression, we cannot have a closed-form solution for $\rho^{*}=\argmax_{\rho \in[0,1] } \widetilde{\eta}(\rho| K) $ directly. However, with the help of existing standard optimization methods, we can still have numerical values for the optimal $\rho$ that maximizes $\widetilde{\eta}(\rho| K) $ . Note that the accuracy of the optimal $\rho$ obtained using the approximated EE is verified in Section~\ref{sec:simulation}.

\section{Simulation Results}\label{sec:simulation}
\label{simulation}

\begin{table}[t]
	\centering
	\caption{Parameter Values}
	\renewcommand{\arraystretch}{1.3}
	\begin{tabular}{c|c}
		\firsthline
		\textbf{Parameters}              & \textbf{Values}  \\
		\hline   
		SBS density ($\lambda_b$) & $10^{-4} /\text{m}^{2}$      \\
		Half cluster center distance ($R_{\text{h}}$)     & 100 m\\
		Pathloss exponent ($\alpha$)     & 4     \\  
		SBS transmit power ($P_{t}$) & 1 W      \\
		Backhaul power per request per SBS ($P_{b}$)          & 10 W    \\
		Available bandwidth ($W$)          & 10 MHz    \\
		SBS cache capacity ($M$)          & $5000$     \\
		Content library size ($N$)          & $10^{5}$   \\
		Zipf shape parameter ($\gamma$)     & $\{0.5, 0.9\}$  \\
		Transmission time fraction ($\beta$)     & $\{0.3, 0.95\}$  \\
		\lasthline
	\end{tabular}
\label{system_params}
\end{table}

In this section, we validate the performance analysis of our cooperative caching and transmission design in cluster-centric SCNs using simulations. The performance is compared with that of cases using only MPC and LCD type caching schemes.

Simulations are performed in a square area of $10^{3} \times10^{3}$ m$^2$. The hexagonal cluster of interest has its cluster center at the origin with distance between two cluster centers equal to $2R_{\text{h}}=200$ m. The approximated circle for the cluster area has radius $R=R_{\text{h}} \sqrt{\frac{2\sqrt{3}}{\pi}}\simeq105$ m. SBSs are distributed according to a homogeneous PPP. All the channel fading follows Rayleigh fading with $|h_i|^{2} \sim \exp(1)$. The values of parameters used for simulation are given in Table \ref{system_params}. 
Simulation results are obtained by averaging over $40000$ realizations. 

Remind that we do not consider the case when there is no SBS in a reference cluster. With our network settings, from \eqref{pmf} we have $\mathbb{P}(n=0)=e^{-2\sqrt{3}\lambda_b R_{\text{h}}^2}=0.03$, meaning that only for $3\%$ of realizations we have empty reference cluster. Therefore, 
excluding empty clusters does not have much impact on the overall network performance.

\begin{figure}
	\centering
	\includegraphics[scale=0.45]{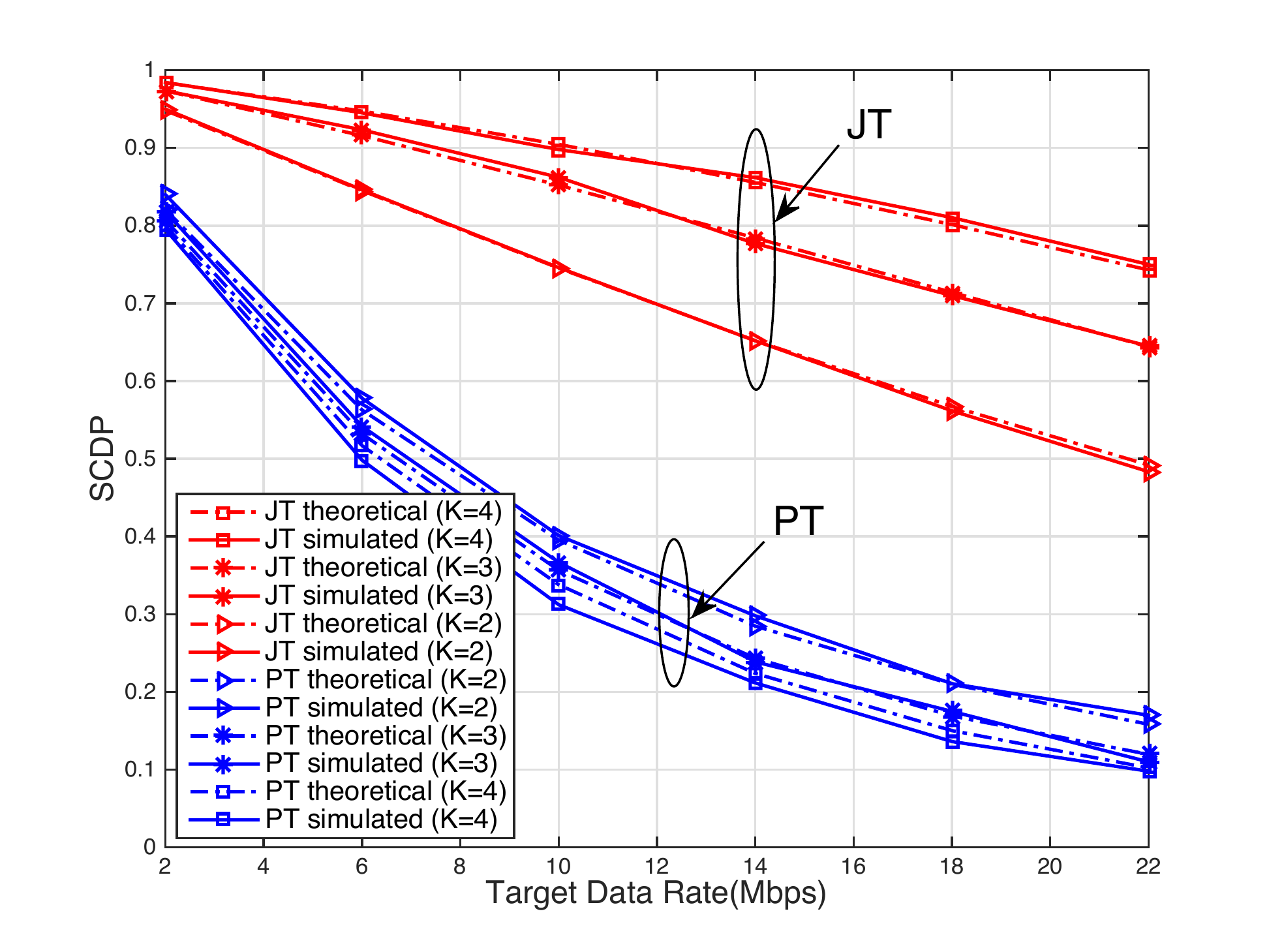}
	\caption{Theoretical and simulation results of the SCDP of JT and PT transmission schemes with $K=\{2, 3, 4\}$.}
	\label{Psuc_simu_nume_K234}
\end{figure}

\subsection{Successful Content Delivery Probability}
Fig. \ref{Psuc_simu_nume_K234} shows the theoretical and simulation results of SCDP of JT and PT (PT-SS) transmission schemes when assuming to have $K=\{2,3,4\}$ SBSs inside the cluster of interest. It first validates the accuracy of our analysis in \eqref{Psuc_JT_expression} and \eqref{Psuc_PT_expression}, especially when $K$ is the close to the average number of SBSs per cluster, i.e., $K=3$. It also proves that the circular approximation of the cluster area has negligible impact on the SCDP analysis. We notice that the error gap in the PT case becomes slightly larger when the conditioned number $K$ is further from the average value $\mathbb{E}[K]=2\sqrt{3}\lambda_b R_{\text{h}}^2$. This is mainly due to the PPP approximation that we use for the interference distribution in \eqref{PTSS_approximation}. When the density of SBSs inside the cluster conditioning on having $K$ SBSs is comparable to the density of PPP distributed out-of-cluster SBSs, the approximation in \eqref{PTSS_approximation} is reasonable. Otherwise the mismatch between the conditioned SBS density inside the cluster and the density of out-of-cluster SBSs causes approximation error in the SIR analysis.  

We also observe that, in the JT case,  higher K yields higher SCDP, but for PT cases, SCDP is lower when K is larger. This is because in the JT case, more cooperative SBSs gives stronger received signal, thus higher SIR value. In the PT case, the SCDP is defined as the product of success probability of multiple streams. When the number of parallel transmitting streams increases, the SCDP will be relatively lower.  

\subsection{Optimization Study of the Combined Caching Strategy}
In the cluster-centric network,  each cluster makes caching decisions independently based on its knowledge about network status inside and outside the cluster, so the optimal percentage  for MPC caching, $\rho^{*}$, is computed in each cluster according to the number of cooperative SBSs $K$.
In this section all the theoretical and simulation results are obtained conditioning on having a certain number $K$ of SBSs inside the cluster of interest.
 
\subsubsection{Cache Service Probability Maximization}
\begin{figure}
	\centering
	\includegraphics[scale=0.45]{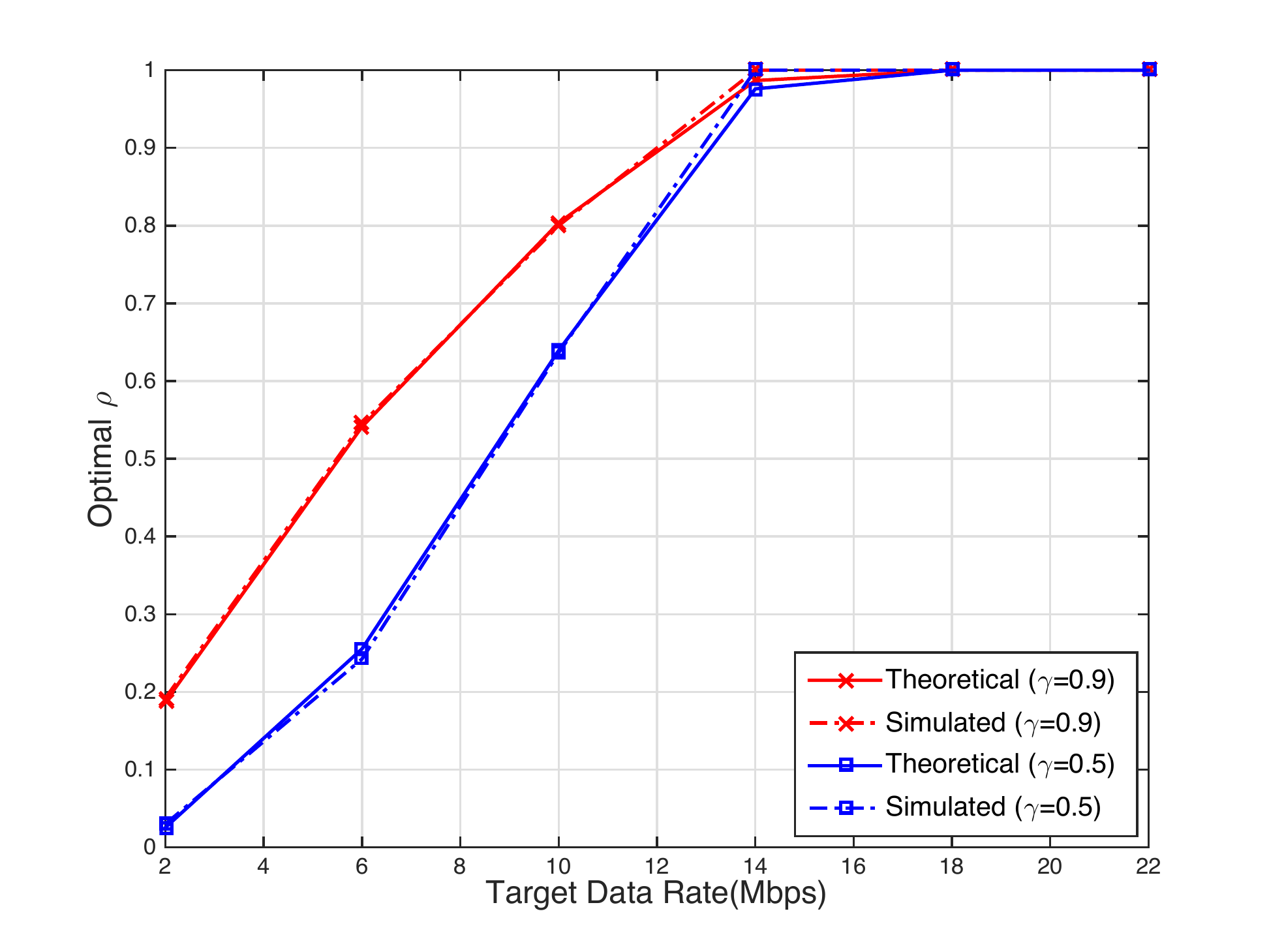}
	\caption{Optimal percentage of MPC type caching, $\rho^{*}$, obtained by cache service probability maximization, when using the proposed combined caching scheme. Both theoretical and simulation results are obtained with $K=3$ and $\gamma=\{0.5, 0.9\}$.}
	\label{optimal_rho_vs_rate}
\end{figure}

In Fig. \ref{optimal_rho_vs_rate}, we plot the optimal $\rho$ obtained in \eqref{optimal_rho}, which maximizes the cache service probability, as a function of the target data rate. The number of in-cluster SBSs is chosen as $K=3$. 
The theoretical optimal values are compared with the real optimum values obtained from the exhaustive search of $\rho$ that maximizes the cache service probability defined in \eqref{objectif_function}. 
We see that $\rho^{*}$ in \eqref{optimal_rho} gives accurate estimation of the real optimum result. We can also see that as expected, $\rho^{*}$ increases with the target rate, because for higher SIR requirement, the transmission reliability is more important for the cache service performance, thus MPC type caching is more favorable. 
The content popularity skewness also affects the optimal MPC cache percentage. When the content popularity is more concentrated, i.e., $\gamma=0.9$, the potential benefit from caching more different files is limited, and in this case, the optimal $\rho$ is expected to be higher, as also shown in Fig.~\ref{optimal_rho_vs_rate}.

\begin{figure}
	\centering
	\includegraphics[scale=0.45]{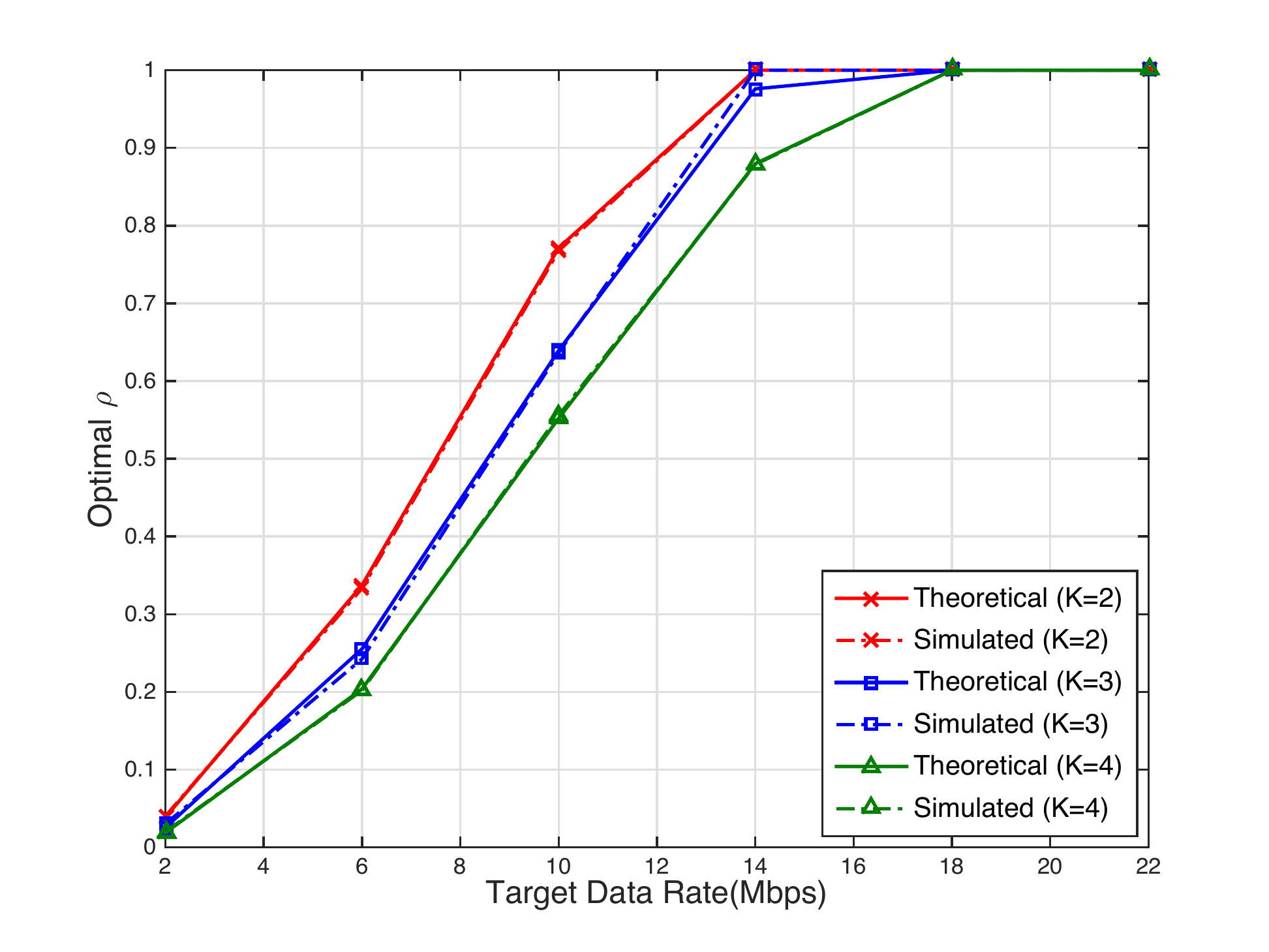}
	\caption{Optimal percentage of MPC type caching, $\rho^{*}$, obtained by cache service probability maximization, when using the proposed combined caching scheme. The results are obtained and presented with $K=\{2,3,4\}$ and $\gamma=0.5$.}
	\label{optimal_rho_diff_K}
\end{figure}

In Fig. \ref{optimal_rho_diff_K}, we plot the theoretical and simulated values of the optimal  $\rho$ conditioning on having $K=\{2, 3, 4\}$ SBSs inside the cluster of interest. The results are obtained with $\gamma=0.5$. Beside the accuracy of the theoretical results, we also notice that for larger $K$, the optimal MPC cache percentage, $\rho^{*}$, is smaller. It shows the potential of improved cooperation gain by reserving more cache space for partition-based LCD caching when the number of cooperative SBSs is larger.

\begin{figure}
	\centering
	\includegraphics[scale=0.45]{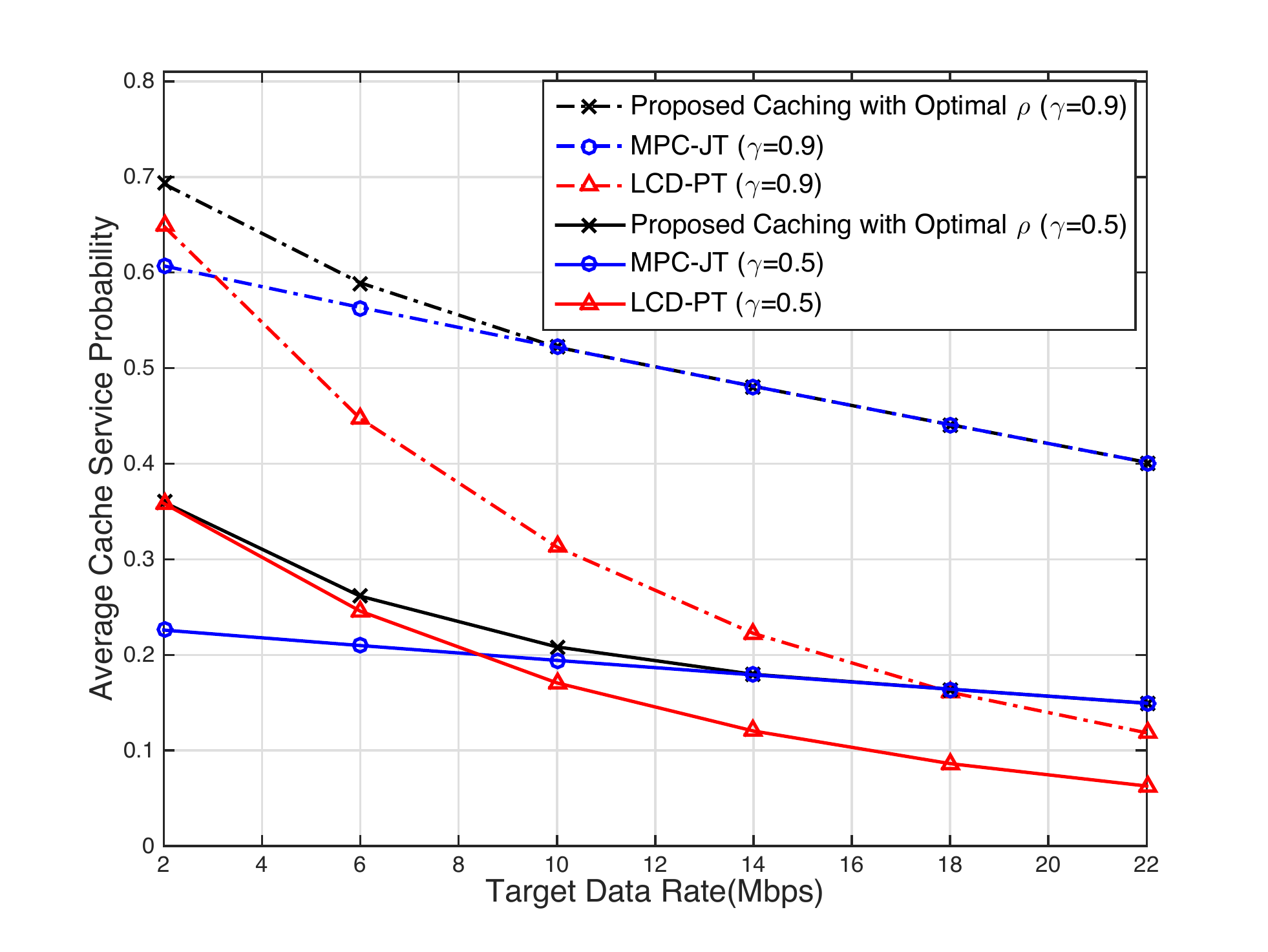}
	\caption{Cache service performance of the proposed combined caching scheme with $\rho^{*}$ given in \eqref{optimal_rho}, with comparison to the case where only MPC or LCD caching is applied. The results are obtained with $\gamma=\{0.5, 0.9\}$ }
	\label{optimal_rho_K3_R100}
\end{figure}

Fig. \ref{optimal_rho_K3_R100} shows the average cache service probability of our proposed cooperative caching and transmission design. The results are obtained by averaging over different values of $K$, as given in \eqref{average_psv}. For practical reasons, we choose $K\in[1,10]$ to get numerical results in the finite range. We see that our proposed caching scheme with optimal $\rho$ derived in \eqref{optimal_rho} always gives better performance than the cases when only either MPC or LCD scheme is applied. As expected, the performance of the proposed caching scheme converges to the performance of LCD and MPC schemes in the extreme cases. 

\subsubsection{Energy Efficiency Maximization}
\begin{figure}
	\centering
	\includegraphics[scale=0.45]{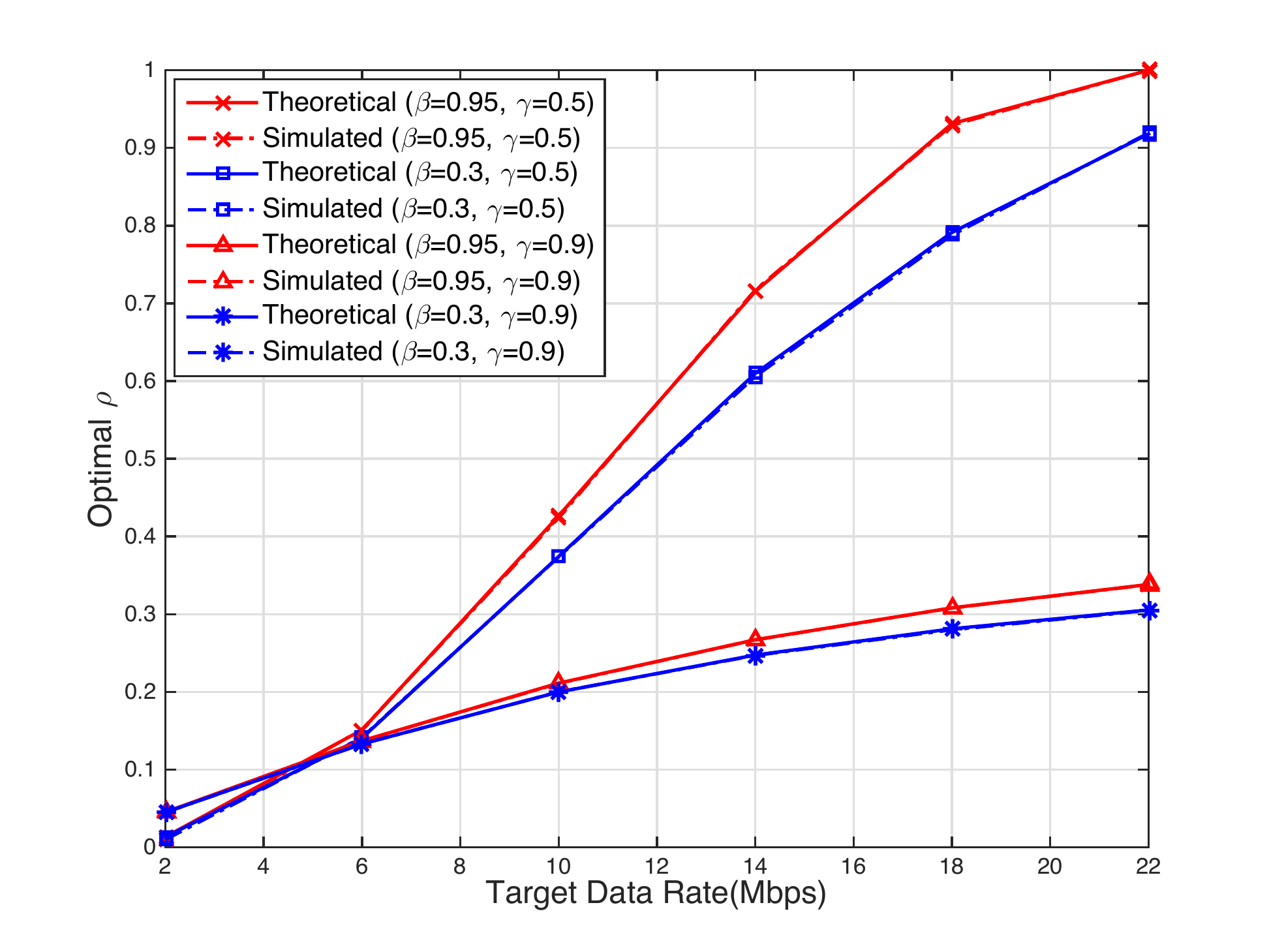}
	\caption{Optimal percentage of MPC type caching, $\rho^{*}$, obtained by the EE maximization when using the proposed combined caching scheme. The results are obtained with $\beta=\{0.95, 0.3\}$, representing the cases with very low and high backhaul delay, respectively, $K=3$, and $\gamma=\{0.5, 0.9\}$.}
	\label{optimal_rho_ee_diff_beta}
\end{figure}

In Fig. \ref{optimal_rho_ee_diff_beta}, we plot the optimal $\rho$ obtained by the EE maximization for different values of backhaul delay. The number of in-cluster SBSs is chosen as $K=3$. The theoretical results are obtained by numerical evaluation of $\rho^{*}=\argmax_{\rho \in[0,1] } \widetilde{\eta}(\rho| K)$ with $\widetilde{\eta}(\rho| K)$ given in \eqref{EE_approximate}. The real optimal values which maximize the EE defined in \eqref{EE_definition} are obtained in simulations by exhaustive searching. We can see that the theoretical $\rho^{*}$ matches well the result obtained in simulation, validating the accuracy of EE maximization with the approximated expression. We also observe the same trend of $\rho^{*}$ as in Fig. \ref{optimal_rho_vs_rate}. When the SIR target increases, the optimal value of $\rho$ is higher, meaning that more space will be assigned for MPC caching. In terms of the impact of backhaul delay on the value of $\rho^{*}$, we see that for higher backhaul delay, i.e., $\beta=0.3$, $\rho^{*}$ is lower, meaning that more space will be assigned for LCD caching in order to avoid fetching the requested content through the backhaul. Compared to the case with cache service probability maximization, $\rho^{*}$ in Fig. \ref{optimal_rho_ee_diff_beta} is always smaller than the ones in Fig. \ref{optimal_rho_vs_rate}, especially in the case with $\gamma=0.9$. We also observe that, when the target rate is relatively high, the optimal $\rho$ obtained with higher $\gamma$ is much smaller than the one obtained with lower $\gamma$. This is because when the popularity is highly concentrated, i.e., $\gamma=0.9$, the benefit of having more space for MPC caching in terms of average rate improvement becomes limited by taking into account the growth trend of the power consumption. It shows the necessity of reserving more space for LCD caching when taking into account the backhaul energy consumption and delay, which coincides with the rationale behind caching in SCNs for improved energy efficiency.

\begin{figure}
	\centering
	\includegraphics[scale=0.45]{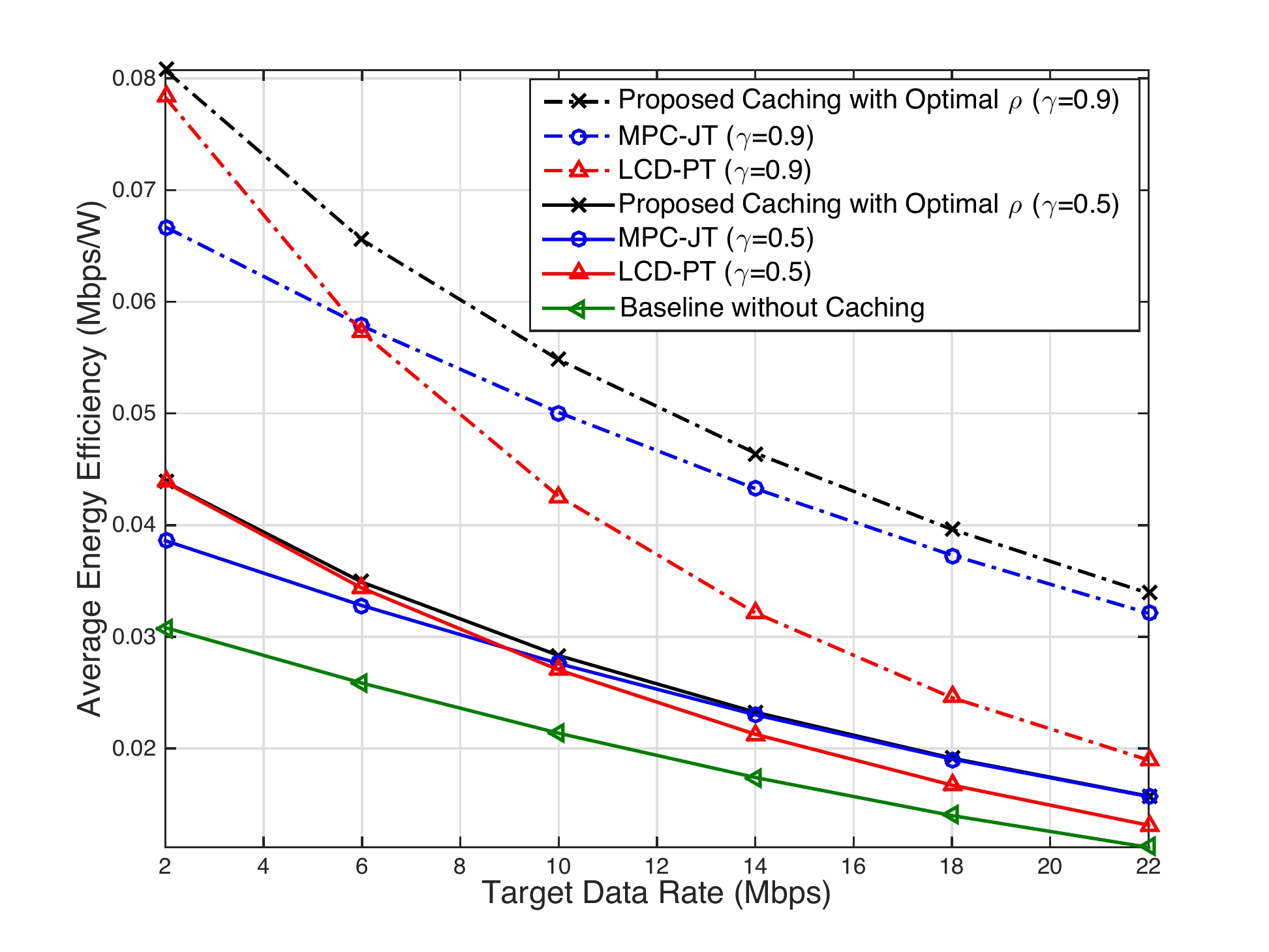}
	\caption{Average energy efficiency of the network when using the proposed combined caching scheme, with comparision to the case with either only MPC or LCD caching. The results are obtained with $\gamma=\{0.5,0.9\}$ and $\beta=0.5$.}
	\label{performance_optimal_rho}
\end{figure}

Fig. \ref{performance_optimal_rho} shows the average EE defined in \eqref{average_ee} when using the optimal $\rho$ obtained by the EE maximization for our proposed caching scheme. The results are compared to the case with only MPC or LCD caching and the baseline result without cache capacity at SBSs. Similar to the case with cache service probability maximization, we observe that our proposed scheme combines the advantage of MPC and LCD caching, thus outperforms the cases where either MPC or LCD caching is applied. Compared to the case without caching, the improvement of EE is validated, showing the benefits of cooperative caching design in SCNs.

An important remark from the presented results on the optimal MPC cache percentage is that, the potential benefit of cooperative caching using PT transmission scheme is strongly limited by the lack of transmission reliability under high data rate requirement. The performance of cooperative caching can be improved by using more advanced SIC techniques, which is out of the scope of this paper, thus will not be discussed here. 

Remind that we evaluate the performance of our proposed cooperative caching and transmission design based on cluster-center user assumption. As presented in Section~\ref{cache_service_optimization}, the optimal MPC cache percentage, $\rho^{*}$, depends on the ratio $\frac{p_{\textnormal{d},K}^{\textnormal{JT}} (\theta_1) }{p_{\textnormal{d},K}^{\textnormal{PT-S}}(\theta_2)}$. The optimal solution for randomly located users in the general case requires analytical results on the SCDP of JT and PT schemes for general users, which are difficult to obtain with neat expressions. The main interest of this work is to introduce the concept and explore benefits of cooperative caching and transmission design in cluster-centric SCNs, therefore, the performance study based on cluster-center user gives sufficient insights on the motivation and design of cooperative SCNs.

\section{Conclusions}
\label{conclusion}
In this work, we studied the potential of using cooperative transmission schemes in cluster-centric cache-enabled SCNs. We proposed a combined MPC and LCD caching strategy with joint and parallel cooperative transmission, respectively. We provided analytical results on the successful content delivery probability of both cooperative transmission schemes for a user located at the cluster center. Our analysis revealed an inherent tradeoff between transmission diversity and content diversity. Motivated by this tradeoff, we solved two optimization problems, namely maximizing the cache service probability and the energy efficiency, respectively. The optimal solutions were given as a function of network parameters and content popularity characteristics. The performance gain of the proposed cooperative caching and transmission design was validated by simulations and compared with simple MPC and LCD caching policies. Our results show that when physical layer cooperation is enabled among SBSs, the performance of content caching can be significantly improved if the caching strategy is duly designed. The main takeaway of this work is that base station cooperation can turn bandwidth into cache memory.

\appendix
\appendices
\subsection{Proof of Lemma \ref{lemma1} }
\label{appen1}
In our network model, we approximately consider the cluster of interest as a circular area  $\mathcal{B}(y_0,R)$, where $y_0$ is the cluster center at the origin. For simplicity, in the following we use $\mathcal{B}(0,R)$ to represent the cluster area. SBSs inside the cluster of interest form the cooperation set, denoted by $\mathcal{C}=\{b_i \in \Phi_b \cap \mathcal{B}(0, R)\}$. Conditioning on having $K$ cooperative SBSs jointly transmitting to the same user,  from the definition of SCDP of JT scheme in \eqref{def_scdp_jt} and the SIR expression in  \eqref{sir_jt}, with target SIR $\theta_1=2^{\frac{R_d}{W}}-1$, we have  
\begin{align}
p_{\textnormal{d},K}^{\text{JT}}(\theta_1) &= \mathbb{P}\left[\left|\sum_{b_i\in\mathcal{C}}h_i {r_i}^{-\frac{\alpha}{2}}\right|^2>\theta_1\!\!\!\!\sum\limits_{b_j\in \Phi_b \backslash \{\mathcal{C}\}} \!\!\!\!|h_j|^2 {r_j}^{-\alpha}\right]  \nonumber\\
&= \mathbb{P}\left[\left|\sum_{i=1}^{K}h_i {r_i}^{-\frac{\alpha}{2}}\right|^2>\theta_1\!\!\!\!\sum\limits_{b_j\in \Phi_b \backslash \mathcal{B}(0,R)} \!\!\!\!|h_j|^2 {r_j}^{-\alpha}\right]. 
\end{align}
Knowing that $\left|\sum\limits_{i=1}^{K}h_i {r_i}^{-\frac{\alpha}{2}}\right|^2 \sim \exp\left(1/\sum\limits_{i=1}^{K}r_i^{-\alpha}\right)$ because of the property of the sum of normally distributed random variables, then we have 
\begin{align}
&p_{\textnormal{d},K}^{\text{JT}}(\theta_1) = \mathbb{E}_{\mathbf{r}}\left[ \mathcal{L}_{I|R} \left(\frac{\theta_1}{\sum_{i=1}^{K}r_i^{-\alpha}}\right) \;\middle|\; \mathbf{r}  \right]   \nonumber\\
&\simeq \int_{\mathbb{R}^K}\mathcal{L}_{I|R} \left(\frac{\theta_1}{\sum_{i=1}^{K}x_i^{-\alpha}}\right) f_{\mathbf{r}}(x_1, \ldots, x_K) \text{d}x_1 \cdots \text{d}x_K,
\label{p_jt}
\end{align} 
where $\mathcal{L}_{I|R}(s)=\mathbb{E}\left[\exp\left(-s \sum\limits_{b_j\in \Phi_b \backslash \mathcal{B}(0,R)} |h_j|^2 {r_j}^{-\alpha}\right)\right]$ is the Laplace transform of interference coming from out-of-cluster SBSs; $f_{\mathbf{r}}(x_1, \ldots, x_K) $ denotes the joint probability density function (pdf) of the distances  $\mathbf{r}=[r_1,\ldots,r_K]$. 

Since $K$ SBSs are independently and uniformly distributed in the cluster approximated by $\mathcal{B}(0, R)$, we have the pdf of the distance $r_i$ from the $i$-th SBS to the user at the origin as
\begin{equation}
f_{r_i}(x_i)\simeq \left\{
\begin{array}{rcl}
\frac{2x_i}{R^2} & & 0 \leq x_i \leq R\\
0 &  & x_i > R
\end{array} \right.
\end{equation}
for any $i\in [1, K]$. From the i.i.d. property of BPP, the joint pdf of the link distances $\mathbf{r}=[r_1, \ldots, r_K]$ is 
\begin{equation}
f_{\mathbf{r}}(x_1, \ldots, x_K)\simeq  \prod\limits_{i=1}^{K} \frac{2 x_i}{R^2},
\label{pdf_distance_jt}
\end{equation}
with $0 \leq x_i \leq R$, $\forall i\in [1, K]$.
Then \eqref{p_jt} becomes
\begin{equation}
\begin{split}
&p_{\textnormal{d},K}^{\textnormal{JT}}(\theta_1)  \\
& \simeq \int_{0}^{R}\!\!\cdots\int_{0}^{R} \mathcal{L}_{I|R} \left(\frac{\theta_1}{\sum_{i=1}^{K}x_i^{-\alpha}}\right) \prod\limits_{i=1}^{K} \frac{2 x_i}{R^2} \textnormal{d}x_1 \cdots \textnormal{d}x_K.
\end{split}
\label{Psuc_jt_proof}
\end{equation} 

Out-of-cluster interference comes from PPP distributed interfering SBSs with minimum distance $R$ to the cluster-center user. We have the Laplace transform of interference from SBSs located out of $\mathcal{B}(0, x)$, given by 
\begin{align}
\mathcal{L}_{I|x} (s)&=\mathbb{E}\left[\exp\left(- s \sum\limits_{b_j\in \Phi_b \backslash \mathcal{B}(0, x)}|h_j|^2 {r_j}^{-\alpha}\right)\right] \nonumber \\
&\mathop{=}\limits^{(a)}\exp\left(-2\pi \lambda_b \int_{x}^{\infty} \frac{sv^{-\alpha}}{1+sv^{-\alpha}}v \text{d}v\right) \nonumber \\
&\mathop{=}\limits^{(b)} \exp\left(-\pi \lambda_b s^{\frac{2}{\alpha}} \int_{\frac{x^2}{s^{2/\alpha}}}^{\infty} \frac{1}{1+w^{\frac{2}{\alpha}}} \text{d}w\right).
\label{laplace_out_derivation}
\end{align}
Here, $(a)$ follows from the probability generating functional (PGFL) of PPP, and $(b)$ is obtained by the change of variable $w=\frac{v^2}{s^{2/\alpha}}$. Combining \eqref{Psuc_jt_proof} and \eqref{laplace_out_derivation}, we obtain Lemma \ref{lemma1}.

\subsection{Proof of Lemma \ref{lemma2} }
\label{appen2}
From the definition of SCDP of PT-SS scheme in \eqref{def_scdp_ptss}, success content delivery happens when the $K$ streams after SIC are decodable, i.e., $\text{SIR}_k>\theta_2$ for $k=1, \ldots,K$, where $\theta_2=2^{\frac{R_d}{KW}}-1$ is the target SIR. 
Then we have the SCDP of PT-SS scheme, given by
\begin{align}
p_{\textnormal{d},K}^{\text{PT-S}} (\theta_2)
&=\mathbb{P} [\text{SIR}_i>\theta_2, \ldots, \text{SIR}_K>\theta_2]   \nonumber\\
&=\mathbb{E}_{ \widetilde{\mathbf{r}}}\left[\prod\limits_{k=1}^{K} \mathbb{P}\left[\text{SIR}_k>\theta_2 \right]\,\middle|\ \widetilde{\mathbf{r}}\right].
\label{pt_suc}
\end{align}
Here the link distance vector $\widetilde{\mathbf{r}}=[\widetilde{r}_1, \ldots, \widetilde{r}_K]$ is with increasing distance order, where $\widetilde{r}_k$ denotes the distance from the $k$-th nearest SBS to the cluster-center user. With the approximation of cluster area as a circle $\mathcal{B}(0, R)$, using the results on the distance distribution of BPP distributed points in a circular area \cite{distance_finite_network}, we have the pdf of the distance from the furthest in-cluster SBS to the cluster center given as
\begin{equation}
f_{\widetilde{r}_K}(x_K)\simeq \frac{2K}{x_K} \left(\frac{x_K}{R}\right)^{2K}.
\label{pdf_rK}
\end{equation}
The conditional distribution of the distance $\widetilde{r}_{k-1}$ from the $(k-1)$-th nearest SBS to the cluster center knowing the distance $\widetilde{r}_{k}=x_{k}$ from the $k$-th nearest SBS is given by
\begin{equation}
f_{\widetilde{r}_{k-1}}(x_{k-1} | x_{k})=\frac{2}{x_{k}}\cdot\frac{1}{B(1,k-1)}\left(\frac{x_{k-1}}{x_{k}}\right)^{2(k-1)-1}
\end{equation}
where $B(a,b)$ is the Beta function.
Knowing that $f_{\widetilde{r}_{k} }(x_{k}| x_{k+1}, \ldots,x_K)=f_{\widetilde{r}_{k} }(x_{k} | x_{k+1})$ because of the i.i.d. property of a BPP,
we obtain the joint pdf of the distances from the $k$-th nearest SBS to the cluster center for $k=1,\ldots,K$ given as 
\begin{align}
f_{\widetilde{\mathbf{r}}}(x_1, \ldots, x_K)&= f_{\widetilde{r}_{K}}(x_K)f_{\widetilde{r}_{K-1}}(x_{K-1}|x_K)\cdots f_{\widetilde{r}_{1}}(r_1|r_2)\nonumber \\
&\simeq \prod\limits_{k=1}^{K} \frac{2 k\cdot x_k}{R^2} .
\label{pdf_pt}
\end{align} 

At the $k$-th SIC step with $k\in[1,K-1]$, since we approximately consider the distribution of interfering SBS as a homogeneous PPP, then we have  
\begin{align}
\mathbb{P}\left[\text{SIR}_k>\theta_2 \,\middle|\ \widetilde{r}_k \right] 
\simeq &\mathbb{P}\left[\frac{\left|h_k\right|^2 {\widetilde{r}_k}^{-\alpha}}{\sum\limits_{b_j\in \Phi_b \backslash \mathcal{B}(0, \widetilde{r}_k)}|h_j|^2 {r_j}^{-\alpha}}>\theta_2 \right]\nonumber  \\
=& \mathcal{L}_{I|\widetilde{r}_k }(\theta_2\cdot \widetilde{r}_k ^{\alpha}).
\label{Pcan}
\end{align}  
For the last decoded stream, we have
\begin{align}
\mathbb{P}\left[\text{SIR}_K>\theta_2 \,\middle|\ \widetilde{r}_K\right]\simeq &\mathbb{P}\left[\frac{\left|h_K\right|^2 {\widetilde{r}_K}^{-\alpha}}{\sum\limits_{b_j\in \Phi_b \backslash \mathcal{B}(0, R)}|h_j|^2 {r_j}^{-\alpha}}>\theta_2 \right] \nonumber \\
=& \mathcal{L}_{I|R }(\theta_2\cdot \widetilde{r}_K ^{\alpha}).
\label{Pdec}
\end{align}
Combining \eqref{Pcan} and \eqref{Pdec}  with the joint pdf in \eqref{pdf_pt}, \eqref{pt_suc} becomes
\begin{equation}
\begin{split}
p_{\textnormal{d},K}^{\textnormal{PT-S}}(\theta_2)
\simeq  
& \int\limits_{0<x_1<\cdots<x_K<R} 
\frac{2 K\cdot x_K}{R^2} \mathcal{L}_{I|R}\left(\theta_2 x_K^\alpha\right)\\
& \quad\times \prod\limits_{k=1}^{K-1} \frac{2 k\cdot x_k}{R^2} \mathcal{L}_{I|x_k}\left(\theta_2 x_k^\alpha\right)
\textnormal{d}x_1 \cdots \textnormal{d}x_K ,
\end{split}
\end{equation}
where $\mathcal{L}_{I|x}(s)$ is given in \eqref{laplace_out_derivation}.

\subsection{Proof of Lemma \ref{lemma3} }
\label{appen3}
In the PT-OS case, due to the orthogonal spectrum usage among in-cluster SBSs, interference only comes from out-of-cluster SBSs. Under the circular approximation $\mathcal{B}(0, R)$ of the cluster area, the interfering SBSs have minimum distance $R$ to the cluster-center user.
For each received stream $i$, with target SIR $\theta_1=2^{\frac{R_d}{W}}-1$, we have the CCDF of SIR, given by
\begin{align}
\mathbb{P}\left[\text{SIR}_i>\theta_1 \,\middle|\ r_i\right]&\simeq\mathbb{P}\left[\frac{|h_i|^2 {r_i}^{-\alpha}}{\sum\limits_{b_j\in \Phi_b \backslash \mathcal{B}(0,R)} |h_j|^2 {r_j}^{-\alpha}}>\theta_1\right]  \nonumber  \\
&=\mathcal{L}_{I|R}\left(\theta_1 r_i^{\alpha}\right).
\label{ccdf_psuc_pt}
\end{align}

Since the instantaneous $\text{SIR}_i$ of each stream is independent of each other,  $\min\{\text{SIR}_i\}>\theta_1$ is equivalent to the event that all $K$ streams satisfy $\text{SIR}_i>\theta_1$. With the help of the approximated joint pdf of $\mathbf{r}=[ r_1, \ldots, r_K]$ in \eqref{pdf_distance_jt},  we have the SCDP of the PT-OS case, given as
\begin{align}
& p_{\text{d},K}^{\text{PT-O}} (\theta_1)  = \mathbb{P}\left[\min\limits_{i\in[1,\ldots, K]}\{\text{SIR}_i\}> \theta_1 \right] \nonumber \\
&\simeq\mathbb{E}_{ \mathbf{r}}\left[\prod\limits_{k=1}^{K} \mathbb{P}\left[\text{SIR}_k>\theta_1 \right]\,\middle|\ \mathbf{r}\right]  \nonumber \\
&=  \int_{0}^{R}\cdots\int_{0}^{R}  \prod\limits_{i=1}^{K} \frac{2 x_i}{R^2}. \mathcal{L}_{I|R}\left(\theta_1 x_i^{\alpha}\right)\text{d}x_1 \cdots \text{d}x_K,
\end{align}  	
where $\mathcal{L}_{I|x}(s)$ is given in \eqref{laplace_out_derivation}.

\subsection{Proof of Lemma~\ref{lemma4} }
\label{appen4}
For simplicity, we use $f(\rho)=f(\rho|K)$ when $K\in[2, \infty]$ is a fixed value. We exclude the case with $K=1$ because it does not require any cache space assignment. 
The simplified cache service probability $f(\rho)$ in \eqref{approximate_function} is twice differentiable in $\rho\in[0,1]$. The second order derivative is
\begin{equation}
\begin{split}
f''(\rho)=&\gamma \rho^{-\gamma-1} \left[p_{\textnormal{d},K}^{\text{PT-S}} (\theta_2) -p_{\textnormal{d},K}^{\text{JT}}  (\theta_1)\right]\\
& -p_{\textnormal{d},K}^{\text{PT-S}} \gamma \left[\rho(1-K)+K\right]^{-\gamma-1} (1-K)^2.
\end{split}
\end{equation}
Knowing that  $p_{\textnormal{d},K}^{\text{PT-S}} (\theta_2) < p_{\text{d},K}^{\text{JT}} (\theta_1)$ from the results presented in Section \ref{tradeoff}, $f''(\rho)$ is always negative, thus $f(\rho)$ is strictly concave.

The first order derivative is
\begin{equation}
\begin{split}
f'(\rho)=&(p_{\textnormal{d},K}^{\text{JT}}  (\theta_2) -p_{\textnormal{d},K}^{\text{PT-S}} (\theta_1) ) \rho^{-\gamma} \\
& + p_{\textnormal{d},K}^{\text{PT-S}} (\theta_2)\left(\rho(1-K)+K\right)^{-\gamma} (1-K).
\end{split}
\end{equation}
Here, $f' (0)$ is positive, and we observe followings.
\begin{itemize}
	\item If $f' (1)\geq0$, that is, $\frac{p_{\textnormal{d},K}^{\text{JT}}  (\theta_1)}{p_{\textnormal{d},K}^{\text{PT-S}} (\theta_2) }\geq K$, $f(\rho)$ monotonically increases in $\rho\in[0,1]$, and the optimal solution is $\rho^{*}=1$.
	\item If $f' (1)<0$, that is, $1<\frac{p_{\textnormal{d},K}^{\text{JT}}  (\theta_1)}{p_{\textnormal{d},K}^{\text{PT-S}}  (\theta_2)} <K$, the optimal solution is the one that satisfies $f'(\rho)=0$. Then, we have $\rho^{*}\simeq  K\left[\left(\frac{K-1}{\frac{p_{\text{d},K}^{\text{JT}} (\theta_1) }{p_{\text{d},K}^{\text{PT-S}} (\theta_2) }-1}\right)^{1/\gamma}+K-1\right]^{-1}
	\label{optimal_percentage}$.
\end{itemize}
By combining both cases together, we get \eqref{optimal_rho} in Lemma \ref{lemma4}.

 
\bibliographystyle{IEEEtran}

\bibliography{bib_ref}

\end{document}